\documentclass[a4paper,11pt]{article}
\pdfoutput=1 % if your are submitting a pdflatex (i.e. if you have
\usepackage{jheppub} % for details on the use of the package, please
\usepackage{bm}
\usepackage[T1]{fontenc} % if needed
\usepackage{amsfonts,amssymb,amsmath}
\usepackage{xcolor, comment,color}
\usepackage{stmaryrd}

\newcommand\abs[1]{\left|#1\right|}

\DeclareMathOperator{\sech}{sech}

\usepackage{ulem}
\title{\boldmath Collision, mechanism, and $Z_4$ operation among chiral and nonchiral kinks in coupled double-field $\phi^4$ model}

\author[a]{Jin-Hyung Choi}
\author[a]{Sang-Hoon Han}
\author[a]{Myungjun Kang}
\author[a, b, 1]{Sangmo Cheon\note{Corresponding author.}}
\affiliation[a]{Department of Physics, Hanyang University, Seoul 04763, Korea}
\affiliation[b]{Research Institute for Natural Science, Hanyang University, Seoul 04763, Korea}

\emailAdd{jhchoihanyang@gmail.com}
\emailAdd{oksk0729@gmail.com}
\emailAdd{mjunkang@gmail.com}
\emailAdd{sangmocheon@hanyang.ac.kr}

\abstract{

In this work, we investigate collision processes and their mechanism among chiral and nonchiral kinks in the coupled double-field $\phi^4$ model, and show that the kink collisions follow the $Z_4$ abelian group operation.
Unlike the single-field $\phi^4$ model, this model has twelve kinks, which are classified into chiral and nonchiral kinks depending on their topological chiral charges. 
This enriches the variety of the collision processes.
From the numerical simulation, we observe three kinds of collisions depending on the initial configuration and initial velocities of colliding kinks.
During a collision, the topological chiral charges of kinks switch while preserving the $Z_4$ abelian group operation.
To understand the collision and chirality switching mechanism, we investigate the detailed collision process, energy densities, the field gradients, internal modes, energy exchange between two fields, coherent vibration of two bions located in different fields, and the orbits of colliding kinks in the two-dimensional field space.
}

\begin{document} 
\maketitle
\flushbottom

\section{Introduction}
% General Introduction
A topological soliton (or kink) is a particle-like localized wave packet, which has attracted wide attention in various areas of modern science such as high energy physics, cosmology, condensed matter, optics, and biological systems \cite{manton2004topological, rajaraman1982solitons,dauxois2006physics,vachaspati1990formation,vilenkin1994cosmic,akhmediev2008dissipative}.
In a strict mathematical sense, a topological soliton maintains its original shape and speed after collisions in an integrable system.
In a broad sense, inelastic scatterings between solitons are also allowed in a nonintegral system, which enriches the variety of soliton dynamics \cite{dauxois2006physics}.
The topological soliton is classified according to mappings from the points at spatial infinity to the degenerate vacua of fields (or homotopy group).
% Such topological classification is applied to various solitons such as topological defect, vortex, magnetic monopole, skyrmion, and so on. 
Hence, a natural question is what happens to the solitons and their topological charges when solitons collide.

Topological soliton dynamics in (1+1)-dimensional field theories have been actively studied \cite{ shnir2018topological,cuevas2014sine,hirota1972exact,ablowitz1973method,sugiyama1979kink, campbell1983resonance, campbell1986solitary, goodman2005kink, goodman2007chaotic, simas2016suppression, peyrard1983kink, dorey2011kink, gani2014kink, weigel2014kink, demirkaya2017kink, marjaneh2017multi, saxena2019higher, gani1999kink, popov2005perturbation, gani2018scattering, gani2019multi, mohammadi2019affective, simas2020solitary, halavanau2012resonance, riazi2002soliton, alonso2018kink, bazeia1997soliton, bazeia1995solitons, bazeia1996solitons, alonso2013new, aguirre2020extended, riazi2012families, nitta2015non, alonso2018reflection, alonso2020non}. %ref
In an integrable system such as the sine-Gordon model, an elastic soliton scattering with a slight phase shift in a soliton's position is studied via inverse scattering method \cite{ablowitz1973method,cuevas2014sine}.
On the other hand, in a nonintegrable system such as the single-field $\phi^4$ model, the kink collision shows richer phenomena such as the formation of the bound state (or bion), multiple bounce resonance, and fractal resonance structure, %depending on the initial velocity of a colliding kink, 
which are investigated via numerical and collective-coordinate methods \cite{sugiyama1979kink,campbell1983resonance, campbell1986solitary, goodman2005kink, goodman2007chaotic, simas2016suppression}.
These exotic features are attributed to the excitation of internal modes and explained by the resonant energy exchange between translational (or zero) and vibrational modes.
Furthermore, kinks and their dynamics have been studied in various nonintegrable single-field models \cite{peyrard1983kink, dorey2011kink, gani2014kink, weigel2014kink, demirkaya2017kink, marjaneh2017multi, saxena2019higher, gani1999kink, popov2005perturbation, gani2018scattering, gani2019multi, mohammadi2019affective, simas2020solitary} and multifield models \cite{halavanau2012resonance, riazi2002soliton, alonso2018kink, bazeia1997soliton, bazeia1995solitons, bazeia1996solitons, alonso2013new, aguirre2020extended, riazi2012families, nitta2015non, alonso2018reflection, alonso2020non}.
In particular, the multifield models show complex and plentiful kink dynamics due to the additional degree of freedom.
However, due to the complexity of the kink collision, the collision mechanism is not clearly understood, and it is necessary to develop a proper tool such as the collective-coordinate method \cite{sugiyama1979kink,takyi2016collective} for multifield models \cite{rajaraman1982solitons}.

Topological solitons possessing chirality (right- or left-handedness) have been realized as hallmarks of various 2D and 3D topological systems:
Helical surface states in topological insulators \cite{zhang2009topological,hasan2010,franz2013topological}, Majorana zero modes in chiral topological superconductors \cite{qi2010chiral, qin2019chiral, elliott2015colloquium,sato2017topological}, chiral edge modes of light in photonic crystals \cite{gao2015topological, ozawa2019topological,barik2020chiral}, and skyrmions in chiral magnets \cite{muhlbauer2009skyrmion,komineas2015skyrmion,tikhonov2020controllable,rho2016multifaceted,zhang2018chiral}.
These topological objects are expected to be used in future information technology such as quantum computation. 
For the 1D case, three types of chiral solitons were realized in the indium atomic wire, and chirality switching between chiral and nonchiral solitons via collision was also demonstrated in the same physical system \cite{cheon2015chiral,kim2017switching}.
This system is described by the double Peierls chain model, which corresponds to the coupled Jackiw-Rebbi field theory composed of two fermion fields and two boson fields \cite{jackiw_solitons_1976, jackiw_solitons_1981,cheon2015chiral,oh2021particle,han2020topological}.
Before these experiments, there were a few theoretical works \cite{bazeia1995solitons, bazeia1996solitons, bazeia1997soliton, riazi2002soliton, halavanau2012resonance, alonso2018kink} about the coupled two field models.
However, studies that reveal the collision dynamics and mechanism among chiral and nonchiral solitons based on chirality are rare.

In this work, we investigate collision processes and their mechanism among chiral and nonchiral kinks in the coupled double-field $\phi^4$ model, and we also show that the topological chiral charges of colliding kinks satisfy the $Z_4$ abelian group operation.
The coupled double-field $\phi^4$ model is composed of two fields ($\phi_1$ and $\phi_2$), a $\phi^4$-type self-interacting potential for each field, and interfield interaction.
The model possesses four degenerate vacua and hence twelve kinks, which are classified into three types depending on the topological chiral charge: right-chiral (RC), left-chiral (LC), and nonchiral (NC) kinks as shown in figure~\ref{fig2:kinks}.
The various chiralities of kinks make the collision scenarios richer while reducing the complexity in understanding kink collision.

All possible collisions between two kinks are numerically studied and classified according to the chirality, which are summarized in table~\ref{table1:collision-table}.
We observe various inelastic collision processes depending on the initial configuration and velocities of colliding kinks.
When two kinks collide, they become trapped and form bions or a kink with a different chirality to the original with a velocity smaller than a critical value.
When it exceeds the critical velocity, the two kinks scatter and escape to infinities.
At some particular velocities, exceptional scatterings occur, which 
is characterized by multiple energy exchange between the two fields. 
Interestingly, in the RC-LC kink collision, we find a resonance between two bions located in different fields.
To understand the collision mechanism, we consider the internal modes, gradient energy density of each field, and energy transfer between two fields. 
Moreover, we investigate the time-dependent orbits of colliding kinks in two-dimensional field space,
which shows the chirality change of kinks explicitly and dynamically.
Because three types of kinks carry $Z_4$ topological chiral charges, all two-kink collisions complete the $Z_4$ abelian group operation table shown in figure~\ref{fig2:kinks}.
Therefore, we believe this work will be helpful for researchers interested in the dynamics of topological solitons with chirality in multifield systems.

\section{System}
\subsection{Double-field $\phi^4$ model\label{sec:double-field-model}}
As a basic building block, we consider the single-field $\phi^4$ model \cite{sugiyama1979kink,campbell1983resonance,campbell1986solitary,goodman2005kink,goodman2007chaotic} in $(1+1)$D, which is described by the following Lagrangian density:
\begin{align}
\label{eq:2.1:Lag}
    \mathcal{L} =
    \frac{1}{2}  \partial_{\mu} \phi  \partial^{\mu} \phi - V(\phi),
\end{align}
where $\phi(t,x)$ is a real scalar field.
The self-interacting potential $V(\phi)$ is given by
\begin{align}
\label{eq:2.2}
    V(\phi) = - \frac{m^2}{2} \phi^2 + \frac{\lambda}{4} \phi^4 + \frac{m^4}{4 \lambda},
\end{align}
where $m$ and $\lambda$ are positive intra-field coupling constants.
The field equation is given by
\begin{align}
    \label{eq:2.3:single field equation}
    \partial_t^2 \phi-\partial_x^2 \phi - m^2 \phi + \lambda \phi^3  = 0.
\end{align}
This model has two degenerate vacua at $\mu_0 = \sqrt{\frac{m^2}{\lambda}} $ and 
the Lagrangian has a discrete $Z_2$ symmetry under  $\phi(x) \rightarrow - \phi(x)$.
The dynamics and mechanism of kink-antikink collisions in the single-field $\phi^4$ model are studied by various methods \cite{sugiyama1979kink,campbell1983resonance, campbell1986solitary, goodman2005kink, goodman2007chaotic, simas2016suppression}.

\begin{figure}[tb]
\centering
\includegraphics[width=.8\textwidth,origin=c]{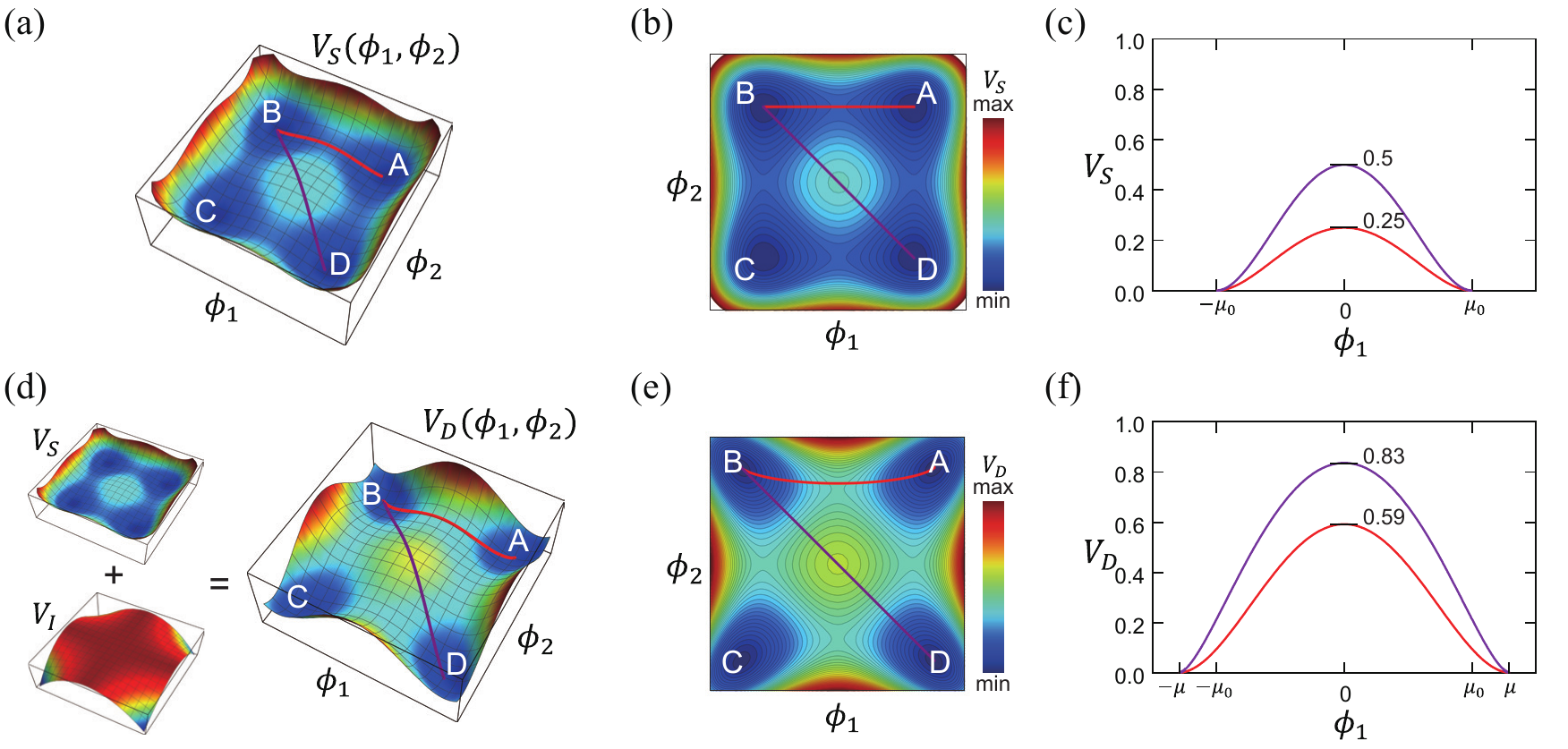}
\caption{
    \label{fig1:model} 
    \textbf{Potentials for the double-field $\phi^4$ models without and with inter-field couplings.}
    \textbf{(a,b)} [\textbf{(d,e)}] 3D and contour plots of the potentials  with respect to $\phi_1$ and $\phi_2$ without (with) inter-field coupling.
    $V_S$ and $V_{I}$ are the self-interacting and inter-field potentials, respectively.
    \textbf{(c)} [\textbf{(f)}] Potentials along the two representative red and purple orbits
    in (a) [(d)].
    $\alpha=0$ for (a-c) and $\alpha=-0.4$ for (d-f).
    The red and purple orbits corresponds to chiral and non-chiral kinks in figure~\ref{fig2:kinks}, respectively.
    }
\end{figure}

We now consider the coupled double-field $\phi^4$ model as a minimally extended model from the single-field $\phi^4$ model.
\begin{align}
\label{eq:2.4;LD}
     \mathcal{L}_{D} =
     \frac{1}{2}  \partial_{\mu} \vec{\phi} \cdot \partial^{\mu} \vec{\phi} 
     - V_{D}( \vec{\phi} ),
\end{align}
where $\vec{\phi}(t, x)= (\phi_{1}(t, x), \phi_{2}(t, x) )^T $, $\phi_i(t,x)$ is a real scalar field, and $V_{D}(\vec{\phi}) = V_{S}(\vec{\phi}) + V_{I}(\vec{\phi})$ is a double-field potential.
Here, $V_{S}(\vec{\phi})$ and $V_{I}(\vec{\phi})$ are self-interacting  and interfield potentials, respectively:
\begin{align}
\label{eq:2.5:V0}
    V_{S}(\vec{\phi}) = V (\phi_1) + V (\phi_2),~~~~~~
    V_{I}(\vec{\phi}) = \frac{\alpha}{2}\phi_1^2\phi_2^2,
\end{align}
where $\alpha$ is an interfield coupling constant.
For an attractive (repulsive) interfield potential,  $\alpha<0$ ($\alpha>0$).
In this work, we only consider the attractive potential because both chiral and nonchiral kinks are stable under perturbations for the attractive potential only, which will be discussed in section~\ref{sec:2.2.Kinks}.
The Lagrangian density has four mirror symmetries:
\begin{align}
\label{eq:2.5:symmetry-1}
    \phi_{1} \rightarrow - \phi_{1},~~ 
    \phi_{2} \rightarrow - \phi_{2},~~
    \phi_1  \leftrightarrow \phi_2,~~ 
    \phi_1  \leftrightarrow - \phi_2.
\end{align}
Moreover, the Lagrangian density has $Z_4$ symmetry which is given by
\begin{align}
\label{eq:2.6:symmetry-2}
    (\phi_1, \phi_2)  \rightarrow (\phi_2, -\phi_1).
\end{align}
In this model, there exist four degenerate vacua labeled as $A$, $B$, $C$, and $D$ [see figure~\ref{fig1:model}].
In the absence of the interfield coupling ($\alpha=0$), the four degenerate vacua are located at
$(\phi_1,\phi_2) = (\pm \mu_0 , \pm \mu_0)$ as shown in figure~\ref{fig1:model}(a,b).
For an attractive interfield coupling ($\alpha<0$), the groundstates are located at $(\phi_1,\phi_2) = (\pm \mu, \pm \mu)$, where 
$ \mu = \sqrt{\frac{\lambda}{\lambda+\alpha}}~\mu_0 $ as shown in figure~\ref{fig1:model}(d,e).
We redefine the self-interacting potential $V_{S}(\vec{\phi})$ and the interfield potential $V_{I}(\vec{\phi})$ to be zero at the vacua by adding constants:
\begin{align}
    V_{S}(\vec{\phi}) = \sum_{j=1,2} \left[ \frac{\lambda}{4} \left( \phi_j^2 - \mu_0^2 \right)^2 -\frac{\lambda}{4} \left( \mu^2 - \mu_0^2 \right)^2 \right],
    ~~~~
    V_{I}(\vec{\phi}) = \frac{\alpha}{2} \phi_1^2 \phi_2^2 -\frac{\alpha}{2} \mu^4.
\end{align}
We will use these potentials from now on.
From the Lorentz-invariant Lagrangian density in (\ref{eq:2.4;LD}), the Lorentz covariant field equations for two fields $\phi_j(x)$ are given by
\begin{align}
\label{eq:2.8:EoM}
    &\partial_{t}^2 \phi_j
    -\partial_{x}^2 \phi_j
    +\partial_{\phi_j} V_{D}=0,~~~~~~(j=1,2).
    % \\
    % & \partial_{t}^2 \phi_i
    % -\partial_{x}^2 \phi_i
    % - m^2 \phi_i + \lambda \phi^3 + g \phi_i \phi^2_{3-i}=0,~~~~~~(i=1,2).
\end{align}
%

%%%%%%%%%%%%%%%%%%%%%%%%%%%%%%%%%%%%%%%%%%%%%%%%%%%%%%%%%%%%%%%%%%%%%
%% Figure 01
%%%%%%%%%%%%%%%%%%%%%%%%%%%%%%%%%%%%%%%%%%%%%%%%%%%%%%%%%%%%%%%%%%%%%

The total energy of the system is given by
\begin{align}
\label{eq:2.14:Energy}
    E_{tot}[\vec{\phi}]
    = \int_{-\infty}^{\infty} dx ~ \epsilon_{tot}(t,x)
    = \int_{-\infty}^{\infty} dx
    \left[ 
        \sum_{i=1,2} \left( \frac{1}{2}(\partial_{t} \phi_i)^2 + \frac{1}{2}(\partial_{x} \phi_i)^2 \right)
        + V_{D}(\vec{\phi})
    \right],
\end{align}
where $\epsilon_{tot}(t,x)$ is the total energy density.
Without the time-derivative terms, the static energy $E_{S}$ is defined as
\begin{align}
\label{eq:3.3:static_interfield_energy}
    E_{S}[\vec \phi]
    = \int_{-\infty}^{\infty} dx
    \left[ 
        \sum_{i=1,2} \left(  \frac{1}{2}(\partial_{x} \phi_i)^2 \right)
        + V_{D}(\vec{\phi})
    \right].
\end{align}
To trace the energy exchange between two fields, we define the spatially-integrated energy of each field $E_i[\phi_i]$ (or the $\phi_i$ field energy) and the spatially-integrated interfield energy $E_I$, which are defined as
\begin{align}
\label{eq:field-energy}
    & E_i[\phi_i]
    = \int_{-\infty}^{\infty} dx
    \left[
        \frac{1}{2} (\partial_{t} \phi_i)^2 
        + \frac{1}{2} (\partial_{x} \phi_i)^2
        + V( \phi_i )
    \right],
    \\
    &
    E_{I}[\vec \phi]
    = \int_{-\infty}^{\infty} dx V_{I}(\vec{\phi}).
\end{align}
For convenience, we define kinetic and gradient energy densities for the $j$th field $\phi_j(t,x)$, which are given by
\begin{align}
    \epsilon_{j}^{K}(t,x) = \frac{1}{2} \left( \partial_{t} \phi_j \right)^2,~~~~~
    \epsilon_{j}^{G}(t,x) = \frac{1}{2} \left( \partial_{x} \phi_j \right)^2.
\end{align}

\subsection{Three types of kinks\label{sec:2.2.Kinks}}
\begin{figure}[tbp]
\centering
\includegraphics[width=.8\textwidth,origin=c]{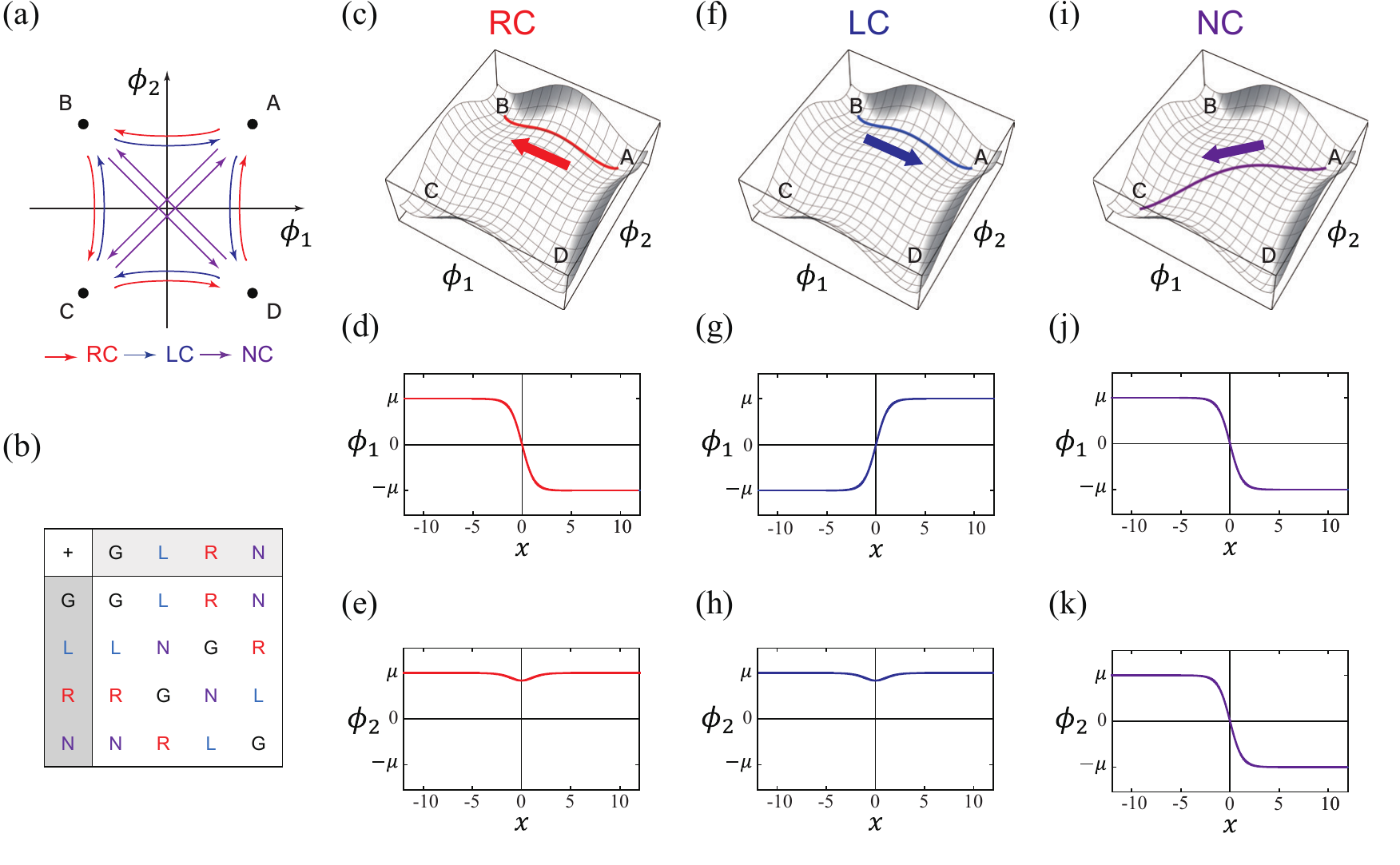}
\caption{
    \label{fig2:kinks} 
    \textbf{Three types of kinks in the coupled double-field $\phi^4$ model.}
    \textbf{(a)}
    Schematic diagram for the four degenerate groundstates $(A, B, C, D)$ and the orbits of twelve kinks in the two-dimensional field space ($\phi_1$, $\phi_2$). 
    The colored arrows represent three types of kinks: RC (right-chiral; red arrow), LC (left-chiral; blue arrow), and NC (non-chiral; violet arrow) kinks.
    \textbf{(b)} Possible $Z_4$ abelian group operation table among groundstates and kinks.
    G, L, R, and N indicate groundstate, LC kink, RC kink, and NC kink, respectively.
    \textbf{(c-k)} 3D potentials and their real-space profiles for (c-e) RC, (f-h) LC, and (i-k) NC kinks.
}
\end{figure}

In this model, there exist kink solutions that interpolate two vacua among four degenerate vacua.
For convenience, a kink is denoted as $\vec{\phi}^{(X,Y)}(t, x)$ and is called as an $XY$ kink when the kink interpolates two distinct vacua $X$ and $Y$ for given time.
Whereas the two degenerate vacua in a single-field $\phi^4$ model permit a pair of kink and anti-kink, the four degenerate vacua in the coupled double-field $\phi^4$ model dictate twelve kinks among them as shown in figure~\ref{fig2:kinks}(a).
Depending on the chirality, these twelve kinks are categorized into three types: right-chiral (RC), left-chiral (LC), and nonchiral (NC) kinks.
The RC (LC) kink is defined as a kink connecting two nearest-neighboring groundstates counterclockwise (clockwise) in two-dimensional field space, while the NC kink interpolates two next-nearest-neighboring groundstates [see figure~\ref{fig2:kinks}(a)]:
\begin{align}
\label{eq:2.15}
    & \text{RC kinks: }
    \vec{\phi}^{(A,B)}(x),~
    \vec{\phi}^{(B,C)}(x),~
    \vec{\phi}^{(C,D)}(x),~
    \vec{\phi}^{(D,A)}(x),
    \\
    & \text{LC kinks: }
    \vec{\phi}^{(A,D)}(x),~
    \vec{\phi}^{(D,C)}(x),~
    \vec{\phi}^{(C,B)}(x),~
    \vec{\phi}^{(B,A)}(x),
    \\
    & \text{NC kinks: }
    \vec{\phi}^{(A,C)}(x),~
    \vec{\phi}^{(C,A)}(x),~
    \vec{\phi}^{(B,D)}(x),~
    \vec{\phi}^{(D,B)}(x).
\end{align}

%%%%%%%%%%%%%%%%%%%%%%%%%%%%%%%%%%%%%%%%%%%%%%%%%%%%%%%%%%%%%%%%%%%%%
%% Figure 02
%%%%%%%%%%%%%%%%%%%%%%%%%%%%%%%%%%%%%%%%%%%%%%%%%%%%%%%%%%%%%%%%%%%%%

Furthermore, the chiral and nonchiral kinks are distinguished by the topological chiral charge $Q$, which  is defined as
\begin{align}
    \label{eq:2.16:topological-chiral-charge}
    Q
    = \int_{-\infty}^{\infty} dx~j^0(x)
    =
   \frac{2}{\pi} ~
   \left[
   \tan^{-1} \left ( \frac{\phi_2(x)}{\phi_1(x)} \right)
   \right]_{-\infty}^{+\infty}.
\end{align}
The corresponding conserved current $j^{\mu}$ satisfying $\partial_{\mu} j^{\mu}=0$ is given by
\begin{align}
   j^{\mu} = \frac{2}{\pi} \epsilon^{\mu \nu} \epsilon_{ij} \frac{\phi_i \partial_{\nu} \phi_j}{ | \vec{\phi}|^2 } 
   =
   \frac{2}{\pi} \epsilon^{\mu \nu} \partial_{\nu} \tan^{-1} \left( \frac{\phi_2}{\phi_1} \right).
\end{align}
Then, kinks with the same chirality possesses the same topological chiral charge.
The topological chiral charge of NC kinks can be represented as either positive or negative, while the topological charge for RC and LC kinks possess opposing signs:
\begin{align}
    Q^{RC} = 1, ~~~Q^{LC} = - 1, ~~~Q^{NC} = \pm 2, ~~\text{(mod 4)}
\end{align}
where the modulo 4 comes from the periodicity of the tangent inverse function in (\ref{eq:2.16:topological-chiral-charge}).

Using the mirror and $Z_4$ symmetry transformations in (\ref{eq:2.5:symmetry-1}) and (\ref{eq:2.6:symmetry-2}), we obtain the relations among kinks as well as their topological chiral charges.
The $Z_4$ rotation in (\ref{eq:2.6:symmetry-2}) transforms a kink into another possessing the same topological chiral charge, which supports the equivalence among the same type of kinks.
On the other hand, mirror operations in (\ref{eq:2.5:symmetry-1}) transform an RC kink into an LC kink and vice versa, while an NC kink maintains its type.

In the viewpoint of the topological chiral charges, RC, LC, and NC kinks, including a vacuum, can carry the quaternary-digit topological information (because a vacuum carries no topological chiral charge). 
Since the total topological chiral charge is conserved during  collisions, collisions among chiral and nonchiral kinks would satisfy the $Z_4$ abelian group operation table shown in figure~\ref{fig2:kinks}(b).
For example, an RC kink can be generated from a collision between LC and NC kinks, or RC and LC kinks can be created pairwise from a vacuum.
Even though such a collision can be allowed by the topological chiral charge conservation,
it is crucial to know whether the collision is permitted dynamically,
which will be examined in section~\ref{sec:two_kink_collisions}.

The stable configurations for the chiral and nonchiral kinks can be obtained by solving the field equation in (\ref{eq:2.8:EoM}).
In the static limit ($\partial_t \phi_i=0$), the field equation in (\ref{eq:2.8:EoM}) becomes
\begin{align}
    \label{eq:2.17:static}
    \partial_{x}^2 \vec{\phi}
    = \nabla_{\phi} V_{D}(\vec{\phi}),
\end{align}
where $\vec \phi = (\phi_1, \phi_2)^{T}$ and $\nabla_{\phi} = (\partial_{\phi_1}, \partial_{\phi_2})$.
This static field equation supports solutions for the chiral and nonchiral kinks, which are plotted in both real and field spaces in figure~\ref{fig2:kinks}.

First, we consider the solution of an NC kink. 
Because two scalar fields $\phi_1(x)$ and $\phi_2(x)$ of an NC kink change their sign at $x=\pm \infty$ such that $\phi_i(-\infty) = -\phi_i(\infty)$ [figure~\ref{fig2:kinks}(j,k)], the orbit of the NC kink in two-dimensional field space $(\phi_1, \phi_2)$ becomes a straight line [figure~\ref{fig2:kinks}(i)].
% For an NC kink, there exists an analytical solution.
Moreover, because  an NC kink satisfy the relation of $\phi_1(t, x) = \pm \phi_2(t, x)$, the field equation (\ref{eq:2.17:static}) becomes
\begin{align}
    \label{eq:2.18:NC field equation}
    \frac{\partial^2 \phi_{i}}{\partial x^2} + m^2 \phi_i - (\lambda+ \alpha) \phi_i^3  = 0,
\end{align}
which is the same field equation of the single-field $\phi^4$ model in (\ref{eq:2.3:single field equation}) except for the coupling constant.
Thus, the analytical solution is given by
\begin{align}
    \label{eq:2.19:NC solution}
    \phi_{i}(x) = \pm \mu \tanh \left( \frac{m}{\sqrt{2}}(x-x_0)\right),
\end{align}
where $x_0$ is the center of an NC kink and the characteristic length ($\sqrt{2}/m$) is constant regardless of the interchain coupling. 
For convenience, we call each solution (or sigmoidal function) located in the field $\phi_i$ as a primary kink ($\mathcal{K}_i$).
Therefore, an NC kink is composed of two primary kinks (NC kink $ = \mathcal{K}_1 + \mathcal{K}_2$).

Next, we consider the solutions of RC and LC kinks.
Because the boundary conditions of RC and LC kinks at infinities $(x=\pm \infty)$ are related by the mirror symmetries in (\ref{eq:2.5:symmetry-1}), the solutions of RC and LC also are related by the same mirror symmetries. For example, the field profiles of the RC and LC kinks shown in figure~\ref{fig2:kinks}(d,e,g,h) are related by the transformation of $(\phi_1, \phi_2 ) \rightarrow (-\phi_1, \phi_2)$.
Unfortunately, when $\alpha \neq 0$, there exists no known analytical solution for RC and LC kinks.
Thus, we numerically obtain the solutions of the RC and LC kinks using the Runge-Kutta method, which are plotted in figure~\ref{fig2:kinks}(c-h).
If a field changes its sign $\phi_i(-\infty) = - \phi_i(\infty)$ as shown in figure~\ref{fig2:kinks}(d,g), then a small (non-topological) lump is induced near the kink's center in the other field $\phi_{3-i}(x)$ as shown in figure~\ref{fig2:kinks}(e,h) due to the interfield interaction.
For convenience, we call the former the primary kink ($\mathcal{K}$) and the latter an induced lump ($\mathcal{I}$). 
Thus, in field space, the orbits of the RC and LC kinks are curved toward the origin [figures~\ref{fig2:kinks}(a) and \ref{fig1:model}(e)].
The numerical solutions can be fitted by the following functions:
\begin{align}
    & \phi_{i}(x) = \pm \mu \tanh \left( \frac{Z_1 m }{\sqrt{2}}(x-x_0)\right),
    \\
    & \phi_{3-i}(x) = \pm \mu  + C_1 \sech^2 \left(  C_2 (x-x_0) \right),
\end{align}
where $Z_{1} = 1 + O(\alpha)$ is the renormalization factor for the inverse of the characteristic length of the primary kink ($\mathcal{K}$). % RC and LC kinks.
$C_1$ and $C_2$ are fitting parameters for the induced lump ($\mathcal{I}$).
In figure~\ref{fig2:kinks} , the numerically calculated values are $Z_1 \approx 1.178$, $C_1 \approx 0.213$, and $C_2 \approx 0.648$.

In the previous two paragraphs, the primary kinks and the induced lumps were introduced for convenience.
However, the shapes of the primary kinks and induced lumps are  not maintained during the collision process in general. 
Nevertheless, we will keep this terminology for simplicity.
For example, when there is a similar form with a sigmoid curve, which tends to be a primary kink, we will call it a primary kink.

Because the solutions of chiral and nonchiral kinks have been obtained, their energies can also be calculated.
By integrating (\ref{eq:2.17:static}), we get 
$\partial_x \vec{\phi} \cdot  \partial_x \vec{\phi} = 2 V_{D}(\vec{\phi})$, 
where the integration constant is zero.
Then, the rest energy of a static kink ($M_0$) is given by
\begin{align}
    M_0
    &=\int_{-\infty}^{\infty}
    \left[ \frac{1}{2} \abs{\partial_x \vec{\phi}}^2 + V_{D}(\vec{\phi}) \right] dx
    = \int_{C}  \left[ 2V_{D}(\vec{\phi} ) \right]^{1/2} d\phi,
\end{align}
where $d\phi\equiv |d\vec{\phi}|$ and $C$ is a curve along which the kink connects two degenerate vacua in two-dimensional field space.

In the absence of the interfield interaction ($\alpha=0$), the two scalar fields are independent.
The rest energy of an NC kink is equal to the sum of those of two chiral kinks: $M_{0}^{NC} = 2 M_{0}^{RC} = 2 M_{0}^{LC}$. The potential of an NC kink is twice that of a chiral kink  as shown in figure~\ref{fig1:model}(c).
However, in the presence of an attractive interfield interaction ($\alpha < 0$), the rest energy of an NC kink is not a simple sum of two chiral kinks.
As shown in figure~\ref{fig1:model}(f), the potential of an NC kink is smaller than twice the potential of a chiral kink due to the attractive interaction.
Thus, the rest energy of an NC kink is smaller than the sum of two chiral kinks, which implies that an NC kink is stable under perturbations.
% ; an NC kink does not decay into RC and LC kinks by a small perturbation.
On the other hand, in the presence of a repulsive interfield interaction ($\alpha > 0$), the rest energy of an NC kink is larger than the sum of two chiral kinks, which implies that an NC kink is unstable and decays into two chiral kinks under perturbations.
For this reason, we only consider the attractive potential ($\alpha<0$) in this work.
From the numerical calculation, the rest energies of RC, LC, and NC kinks are given by $M_{0}^{RC} = M_{0}^{LC} = 1.88$ and $M_{0}^{NC}=3.14$ when $\alpha=-0.4$, $\lambda = 1$, and $\mu_0 = 1$.
Because $2 M_{0}^{RC} = 2 M_{0}^{LC} > M_{0}^{NC}$, two RC or LC kinks prefer to merge into an NC kink from an energy perspective.
% 

% moving kinks
We now discuss the general solutions for the traveling kinks, which can be obtained from the static solutions above using the Lorentz boost.
If $\phi_i(x)$ is a static kink solution, then $\phi_i(\xi)$ is a moving kink solution with a fixed velocity $v$ ($-1<v<1$), where $\xi \equiv \gamma (x-x_0-vt)$ and $\gamma\equiv(1-v^2)^{-1/2}$.
Therefore, the characteristic lengths of moving kinks are decreased and the energy of such a moving kink is given by $M =\gamma M_0$.

%%%%%%%%%%%%%%%%%%%%%%%%%%%%%%%%%%%%%%%%%%%%%%%%%%%%%%%%%%%%%%%%%%%%
% Figure 03
%%%%%%%%%%%%%%%%%%%%%%%%%%%%%%%%%%%%%%%%%%%%%%%%%%%%%%%%%%%%%%%%%%%%
\begin{figure}[tbp]
\centering
\includegraphics[width=.8\textwidth,origin=c]{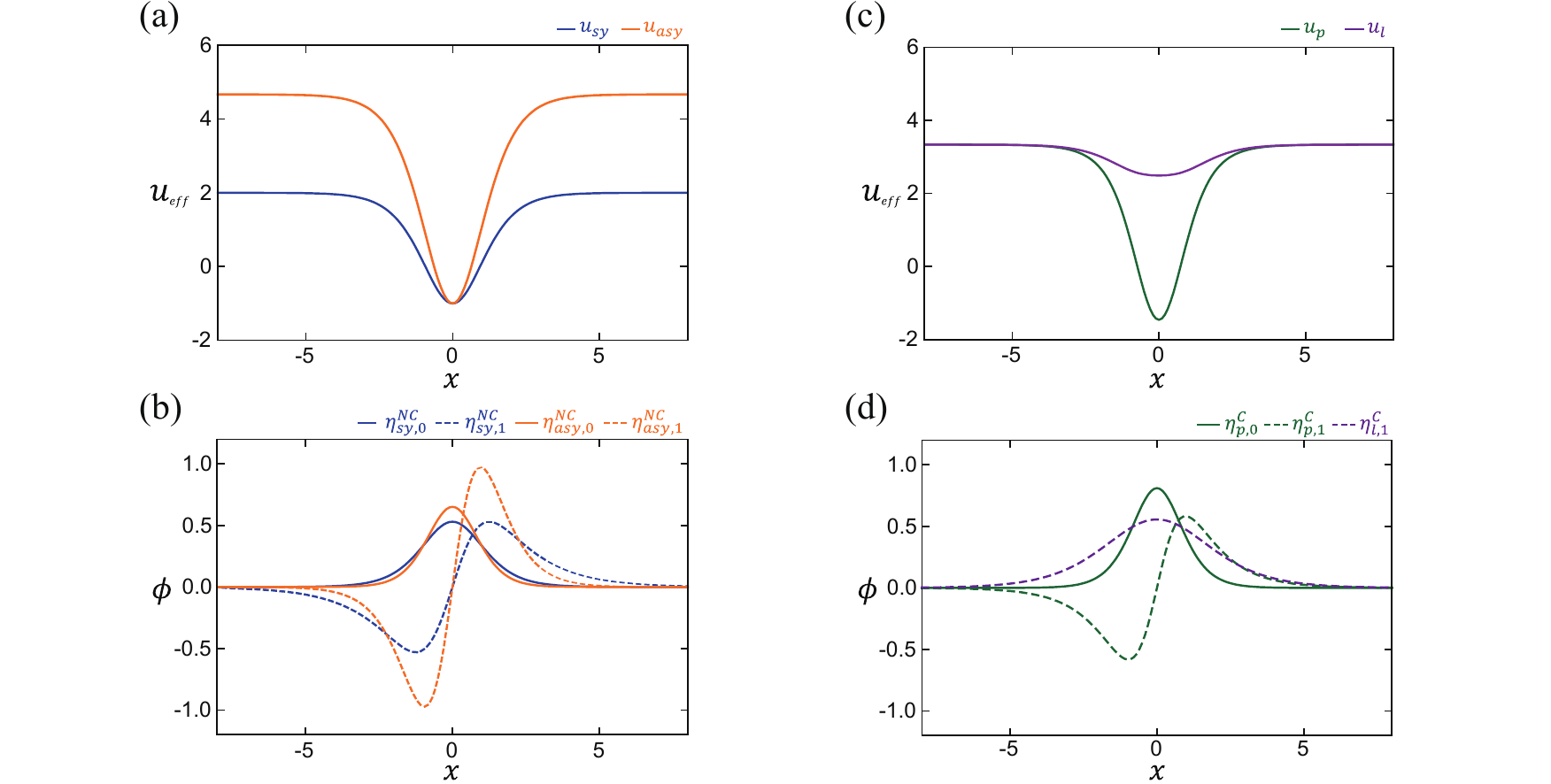}
\caption{
    \label{fig3:internal modes} 
    \textbf{Internal modes for each kinks.}
    \textbf{(a)} Effective potentials $u_{eff}$ for the symmetric (blue) and asymmetric (orange) fluctuations for nonchiral kinks.
    \textbf{(b)}
    $\eta_{sy,0}^{NC}$ and $\eta_{sy,1}^{NC}$ are the internal modes of the symmetric fluctuation,
    while $\eta_{asy,0}^{NC}$ and $\eta_{asy,1}^{NC}$ are of the asymmetric fluctuation.
    \textbf{(c)} Effective potentials for the fluctuations of the primary kink (green) and the induce lump (violet).
    \textbf{(d)} 
    $\eta_{p,0}^{C}$ and $\eta_{p,1}^{C}$ are the internal modes for the primary kink,
    while $\eta_{l,1}^{C}$ is for the induced lump.
}
\end{figure}
\subsection{Internal modes\label{sec:internal modes}}

In this section, we study the oscillating internal modes trapped in each kink by solving the linearized equation \cite{sugiyama1979kink,dauxois2006physics}, the results of which are consistent with ones given by the perturbation method \cite{halavanau2012resonance}.
In the single-field $\phi^4$ model, such oscillating internal modes have been interpreted as low energy excitation trapped in kinks and play an essential role in the inelastic collisions between kinks \cite{sugiyama1979kink,goodman2005kink,goodman2007chaotic,simas2016suppression}.
Similarly, the internal modes in this section is one reason for a collision between kinks being inelastic.

For an NC kink, we consider the following small oscillation around the static solution:
\begin{align}
    \label{eq:2.:NC_fluctuation_0}
    \phi_{i}(t,x) = \mu \tanh \left( \frac{m}{\sqrt{2}} x \right) + \eta_i^{NC}(x)e^{-i \omega t},
\end{align}
where $\eta_i^{NC}(x)e^{-i \omega t}$ is a possible oscillating normal mode for the $i$th field.
By substituting this equation into the field equation (\ref{eq:2.8:EoM}) and ignoring higher terms, we get the following equations for normal modes:
  \begin{align}
    \label{eq:2.:NC_fluctuation_1}
    & \left[  - \frac{\partial^2 }{\partial x^2}
    - m^2 
    + (3\lambda + 3g ) \mu^2 \tanh^2 \left( \frac{m}{\sqrt{2}} x \right)
    \right] \eta_{sy}^{NC}
    = \omega^2 \eta_{sy}^{NC},
    \\
    \label{eq:2.:NC_fluctuation_2}
    & \left[  - \frac{\partial^2 }{\partial x^2}
    - m^2 
    + (3\lambda - g ) \mu^2 \tanh^2 \left( \frac{m}{\sqrt{2}} x \right)
    \right] \eta_{asy}^{NC}
    = \omega^2 \eta_{asy}^{NC},
\end{align}
where $\eta_{sy}^{NC} = \eta_1^{NC}+\eta_2^{NC}$ and $\eta_{asy}^{NC} = \eta_1^{NC}-\eta_2^{NC}$ are symmetric and anti-symmetric modes, respectively.
The eigenvalue equations in (\ref{eq:2.:NC_fluctuation_1}) and (\ref{eq:2.:NC_fluctuation_2}) are formally analogous to time-independent Schr\"odinger equations with the Rosen-Morse type effective potentials \cite{rosen1932vibrations,huang2013solutions},
which give the following four discrete spectra: 
\begin{align}
    \label{eq:2.26:NC_internal_mode_1}
    &\omega_{sy,0}^2=0,
    ~~~ \eta_{sy,0}^{NC}(x) = N_1 \sech^2 \left( \frac{mx}{\sqrt{2}} \right),
    \\
    \label{eq:2.26:NC_internal_mode_2}
    &\omega_{sy,1}^2=\frac{3}{2}m^2,
    ~~~ \eta_{sy,1}^{NC}(x) = N_2 \sinh \left( \frac{mx}{\sqrt{2}} \right) \sech^2 \left( \frac{mx}{\sqrt{2}} \right),
    \\
    \label{eq:2.26:NC_internal_mode_3}
    &\omega_{asy,0}^2=\frac{m^2}{4}
    \left(
        \frac{ \sqrt{ 25 \lambda - 7  \alpha} }{ \sqrt{\lambda+\alpha} } -5 
    \right),
    ~~~ \eta_{asy,0}^{NC}(x) = N_3 
    \left[\sech \left( \frac{mx}{\sqrt{2}} \right) \right]
    ^{\frac{1}{2} \left(\frac{\sqrt{25 \lambda -7 \alpha}}{\sqrt{\lambda +\alpha }}-1\right)},
    \\
    \label{eq:2.26:NC_internal_mode_4}
    &\omega_{asy,1}^2=\omega_{asy,1}^2 + \frac{3}{2}m^2,
    ~~~ \eta_{asy,1}^{NC}(x) = N_4 \sinh \left( \frac{mx}{\sqrt{2}} \right)
    \left[\sech \left( \frac{mx}{\sqrt{2}} \right) \right]
    ^{\frac{1}{2} \left(\frac{\sqrt{25 \lambda -7 \alpha}}{\sqrt{\lambda +\alpha }}-1\right)},
\end{align}
where $N_i$ ($i=1,2,3,4$) are normalization constants.
The solution in (\ref{eq:2.26:NC_internal_mode_1}) is a zero (or translation) mode that recovers the transnational invariance of the model, while the solutions in (\ref{eq:2.26:NC_internal_mode_2})-(\ref{eq:2.26:NC_internal_mode_4}) are excitation modes trapped in the NC kink.
For small interfield coupling ($\alpha \neq 0$), the frequencies of internal modes satisfy the inequality 
$\omega_{sy,0} < \omega_{asy,0} < \omega_{sy,1} < \omega_{asy,1}$.
If $\alpha=0$, the asymmetric modes become the symmetric modes so that $\omega_{asy,0} = \omega_{sy,0} $ and $\omega_{asy,1} = \omega_{sy,1}$, which is consistent with the single-field $\phi^4$ model \cite{sugiyama1979kink}.
The asymptotic form of the wave function $\eta_i^{NC}$ in (\ref{eq:2.:NC_fluctuation_1}) and (\ref{eq:2.:NC_fluctuation_2}) become plane waves at $x= \pm \infty$ satisfying the dispersion relations $\omega_{k_x}^2 = k_x^2 + 2 m^2$ and $\omega_{k_x}^2 = k_x^2 + 2 m^2 (\lambda-\alpha)/(\lambda+\alpha)$, respectively.
Therefore, the continuum spectra are ordinary bosonic modes having masses $\sqrt{2} m$ and  $\sqrt{2(\lambda-\alpha)/(\lambda+\alpha)} m$, respectively, which can be a dissipation channel when a kink collision occurs.
Figure~\ref{fig3:internal modes}(a,b) shows the numerically calculated effective potentials and internal modes.
For chiral kinks, we consider the following small oscillatory modes ($\eta_{p}^{C}$ and $\eta_{l}^{C}$) around the static solution:
\begin{align}
    \label{eq:2.31:RC_internal_mode_1}
    &\phi_{i}(t,x) = \mu \tanh \left( \frac{Z_1 }{\sqrt{2}}x \right) + \eta_{p}^{C}(x) e^{- i \omega t},
    \\
    \label{eq:2.31:RC_internal_mode_2}
    &\phi_{3-i}(t,x) = \mu - C_1 \sech^2 \left(  \frac{C_2 }{\sqrt{2}} x \right) + \eta_{l}^{C}(x) e^{- i \omega t},
    \end{align}
where $\eta_{p}^{C}$ and $\eta_{l}^{C}$ are fluctuations for the primary-kink field ($\phi_{i}$) and the induced-lump field ($\phi_{3-i}$), respectively. 
In the leading order, we study the internal mode by considering the fluctuations separately because the interplay between $\eta_{p}^{C}$ and $\eta_{l}^{C}$ are complex and higher-order (see the detailed coupled equations in appendix~ \ref{app:coupled-equations}).

We first consider the internal modes for the primary kink (the case $\eta_{p}^{C} \neq 0$ and  $\eta_{l}^{C} =  0$).
By substituting (\ref{eq:2.31:RC_internal_mode_1}) and (\ref{eq:2.31:RC_internal_mode_2}) into the field equation (\ref{eq:2.8:EoM}) and ignoring higher order terms, we get the following equation for the normal mode:
    \begin{align}
    \label{eq:2.32:RC_internal_mode_2}
    \left[ - \frac{\partial^2 }{\partial x^2} + u_{\text{eff}}(x) \right] \eta_{p}^{C}(x) = \omega^2 \eta_{p}^{C}(x),
    \end{align}
where $u_{\text{eff}}$ is the effective potential. 
In the limit of small $\alpha$ ($\alpha \ll 1$), the effective potential can be approximated as
    \begin{align}
    u(x) \approx 3 Z_1^2 m^2 \tanh^2 \left( \frac{Z_1 m x}{\sqrt{2}} \right) - Z_1^2 m^2.
    \end{align}
Then, the eigenvalue equation in (\ref{eq:2.32:RC_internal_mode_2}) gives two discrete spectra:
    \begin{align}
    \label{eq:2.23:chiral-zero-mode}
    &\omega_{p,0}^2=0,
    ~~~ \eta_{p,0}^{C}(x) = \sqrt{\frac{3 Z_1 m }{4\sqrt{2}}} \sech^2 \left( \frac{Z_1 m x}{\sqrt{2}} \right),
    \\
    \label{eq:2.23:chiral-1st-mode}
    &\omega_{p,1}^2=\frac{3}{2} Z_1^2 m^2,
    ~~~ \eta_{p,1}^{C}(x) = \sqrt{\frac{3 Z_1 m }{2\sqrt{2}}} \tanh \left( \frac{Z_1 m x}{\sqrt{2}} \right) \sech \left( \frac{Z_1 m x}{\sqrt{2}}  \right),
    \end{align}
where (\ref{eq:2.23:chiral-zero-mode}) is the zero (or translational) mode for the translation invariance and (\ref{eq:2.23:chiral-1st-mode}) is the excitation mode.
% These solutions are zero and excitation modes, respectively.
Moreover, there exist continuum modes satisfying the dispersion relation $\omega_k^2 = k_x^2 + 2 Z_1^2 m^2$. Thus, the continuum spectrum is an ordinary bosonic mode having mass $\sqrt{2}Z_1 m$.
Figure~\ref{fig3:internal modes}(c,d) shows the numerically calculated effective potential and internal modes for the primary kink.
The numerically calculated frequencies for the zero and excitation modes are given by $\omega_{p,0}^2 =0$ and $\omega_{p,1}^2 =2.39$, which are consistent with (\ref{eq:2.23:chiral-zero-mode}) and (\ref{eq:2.23:chiral-1st-mode}).

We now consider the internal modes for the induced lump (the case $\eta_{l}^{C} \neq 0$ and  $\eta_{p}^{C} =  0$).
% We now consider the case $\eta_{l}^{C} \neq 0$ and  $\eta_{p}^{C} =  0$.
%By substituting (\ref{eq:2.31:RC_internal_mode_1}) and (\ref{eq:2.31:RC_internal_mode_2}) into the field equation (\ref{eq:2.8:EoM}) and ignoring higher terms, one can get the similar Schr\"odinger equation with (\ref{eq:2.32:RC_internal_mode_2}).
By substituting (\ref{eq:2.31:RC_internal_mode_1}) and (\ref{eq:2.31:RC_internal_mode_2}) into the field equation (\ref{eq:2.8:EoM}) and ignoring higher order terms, a Schr\"odinger-like equation can be derived from (\ref{eq:2.32:RC_internal_mode_2}).
Since there is no analytical solution, the internal mode can be numerically obtained. There exists one internal mode having a wavefunction of the form $\eta_{l}^{C} = N_0 \sech^2 (Z_2 x)$ with frequency $\omega^2_{l} \approx 2.83$ for $\alpha=-0.4$.
(Here, $N_0 \approx 0.372$ and $Z_2 \approx 0.185 $ are normalization and fitting parameters, respectively.)
Figure~\ref{fig3:internal modes}(c,d) shows the numerically calculated effective potential and the internal mode for the induced lump.

\section{Two-kink collisions and $Z_4$ operation\label{sec:two_kink_collisions}}
In this section, we investigate all possible two-kink collisions by solving the field equation (\ref{eq:2.8:EoM}) numerically, and show that all collisions satisfy the $Z_4$ abelian group operation table in figure~\ref{fig2:kinks}.
The results are summarized in table~\ref{table1:collision-table}.
In the numerical simulation, the initial distances between two kinks are large enough to make sure the overlap of the kinks is small.
The initial momenta of two colliding kinks are adjusted to observe the collision in the center-of-mass frame.
We set the 1500 grid for $x$ and 3500 grid for $t$ with grid spacings $dx=0.05$ and $dt=0.025$, which correspond to the range of $-37.5 \leq x \leq 37.5$ and $0 \leq t \leq 87.5$, respectively.

%%%%%%%%%%%%%%%%%%%% Table %%%%%%%%%%%%%%%%%%%%
\begin{table}[t]
\centering
\begin{tabular}{c|ccc|cc}
\noalign{\smallskip}\noalign{\smallskip}
\hline\hline
Type    & $ v < v_c$ & $v > v_c$ & $v = v_{p}$ & $v_c$ & $v_{p}$\\
\hline
NC $\rightarrow \leftarrow$ NC &   Groundstate  &  $\leftarrow$ NC, NC $\rightarrow$  &          & 0.194 &    
\\
RC $\rightarrow \leftarrow$ RC &   NC  &  $\leftarrow$ LC, LC $\rightarrow$  &          & 0.420&  
\\
LC $\rightarrow \leftarrow$ LC &   NC  &  $\leftarrow$ RC, RC $\rightarrow$  &         & 0.420 &  
\\
\hline
RC $\rightarrow \leftarrow$ LC &  Groundstate   & $\leftarrow$ LC, RC $\rightarrow$  & $\leftarrow$ RC, LC $\rightarrow$  & 0.560 & 0.460       
\\
RC $\rightarrow \leftarrow$ NC &   $\leftarrow$ LC&  $\leftarrow$ NC, RC $\rightarrow$  &$\leftarrow$ RC LC, LC $\rightarrow$& 0.791 & 0.785
\\
LC $\rightarrow \leftarrow $  NC  & $\leftarrow$ RC &  $\leftarrow$ NC, LC $\rightarrow$  &$\leftarrow$ LC  RC, RC$\rightarrow$ & 0.791 & 0.785 
\\
\hline
\hline
\end{tabular}
\caption{
\label{table1:collision-table}
\textbf{Two-kink collisions in the coupled double-field $\phi^4$ model and $Z_4$ operation.}
The first three rows show the collisions between the same types of kinks, while the next three rows indicate ones between different kinks.
The left and right arrows indicate the moving direction of kinks before and after the collision.
$v_c$ and $v_{p}$ are the critical and representative particular velocities, respectively.
The topological chiral charges for RC, LC, and NC kinks are $Q^{\text{RC}}=1$, $Q^{\text{LC}}=-1$, and $Q^{\text{NC}}=2$ (mod 4), respectively. Thus, in the viewpoint of the topological chiral charge, all collisions in this table satisfy the $Z_4$ abelian group operation.
(Because the results of the NC-NC kink collision are the same with the previous results in the single-field $\phi^4$ model \cite{campbell1983resonance, campbell1986solitary, goodman2005kink, goodman2007chaotic}, the particular velocities for $n$-bounce scattering in the NC-NC kink collison are omitted in this table.)
}
\label{table 2}
\end{table}
%%%%%%%%%%%%%%%%%%%% Table %%%%%%%%%%%%%%%%%%%%

To understand the collision processes, we examine real-space profiles ($\phi_i$), field gradients densities ($\epsilon^G_i$), total energy densities ($\epsilon_{tot}$), spatially-integrated energies ($E_{S}, E_i, E_I$) and field-space orbits.
(The detailed definitions are given in section~\ref{sec:double-field-model}.)
%
%%%%%%%%%%%%%%%%%%%%%%%%%%%%%%%%%%%%%%%%%%%%%%%%%%%%
%%%% 충돌 과정을 분석하기 위한 툴 : trajectory %%%%%
%%%%%%%%%%%%%%%%%%%%%%%%%%%%%%%%%%%%%%%%%%%%%%%%%%%%
We plot the real-space profiles $\phi_i(t,x)$ and the gradient energy densities $\epsilon^G_i(t,x)$ of two fields.
From the maximum spatial point of $\epsilon^G_i(t,x)$ for a given time, the center of each kink is obtained. The trajectories are defined by connecting the kink centers.
%%%%%%%%%%%%%%%%%%%%%%%%%%%%%%%%%%%%%%%%%%%%%%%%%%%%%%%%%%%%%%%%%%%
%%%% total energy density and spatially-integrated energies   %%%%%
%%%%%%%%%%%%%%%%%%%%%%%%%%%%%%%%%%%%%%%%%%%%%%%%%%%%%%%%%%%%%%%%%%%
We plot the total energy density $\epsilon_{tot}$ and spatially-integrated energies ($E_{i}$ and $E_{I}$).
Most of the energy is localized in the kinks before and after the collision, which implies that the collision is well defined.
The graphs of $E_{i}$ and $E_{I}$ show the energy of $\phi_i$ and the interfield energy, from which the energy transfer among them can be read.
We compare the interfield energy $E_I$ and static energy $E_{S}$
[see detailed numerical results in figure~\ref{fig17:EI-ES}].
As two colliding kinks approach each other, $E_{S}$ decreases before the collision.
This implies that the two kinks are attractive. Moreover, $E_{I}$ decreases more rapidly than $E_{S}$, which means that the interfield coupling makes two colliding kinks attractive.
In particular, the field-space orbits shows the dynamical motion of the kink in two-dimensional field space, and the orbits can be considered as an ideal elastic and stretchable string with zero equilibrium length under potential $V_D$ in two-dimensional space.
Because the Lagrangian in (\ref{eq:2.4;LD}) has exactly the same form with the Lagrangian for such a string, the equation of motion in (\ref{eq:2.8:EoM}) determines the real-space motion of each field as well as the motion of the field-space orbit. 
% \smd{In this viewpoint}
Thus, we define force $\vec{F}$ and velocity $\vec{V}$ vector fields that act on the field-space orbit:
\begin{align}
    \label{eq:3.4:force-velocity}
    F_i (t,x) \equiv \partial_t^2 \phi_i = \partial_x^2 \phi_i - \partial_{\phi_i} V_D = T_i - P_i, ~~~~
    V_i (t,x)  \equiv \partial_t \phi_i,
\end{align}
where $\vec{F}$ can be decomposed into a string tension $T_i = \partial_x^2 \phi_i$ and potential gradient $P_i = \partial_{\phi_i} V_D $, for convenience.
Note that $T_i$ is a force that depends on the detailed field profile of $\phi_i$, while $P_i$ is a force that depends on only the potential $V_D$.
These force and velocity can be represented by arrows on the orbit in two-dimensional field space such that $F_i[\vec \phi (t,x)] = F_i (t,x)$ and $ V_i [\vec \phi (t,x)] = V_i (t,x)$.
It is noteworthy that even though two field-space orbits of two different collisions are similar, $\vec F$ and $\vec V$ vector fields can be dramatically different, which plays an important role in determining the fate of collisions as discussed below.
(See detailed numerical results in appendix~\ref{app:force-velocity}.)
The motions of field-space orbits, $\vec F$, and $\vec V$  during collision processes can be seen in the supplementary movie online.

%
%%%%%%%%%%%%%%%%%%%%%%%%%%%%%%%%%%%%%%%%%%%%%%%%%%%%%%%%%%%%%%%%%%%%%
%% Figure 04 LC-LC main
%%%%%%%%%%%%%%%%%%%%%%%%%%%%%%%%%%%%%%%%%%%%%%%%%%%%%%%%%%%%%%%%%%%%%
\begin{figure}[t]
\centering
\includegraphics[width=.8\textwidth,origin=c]{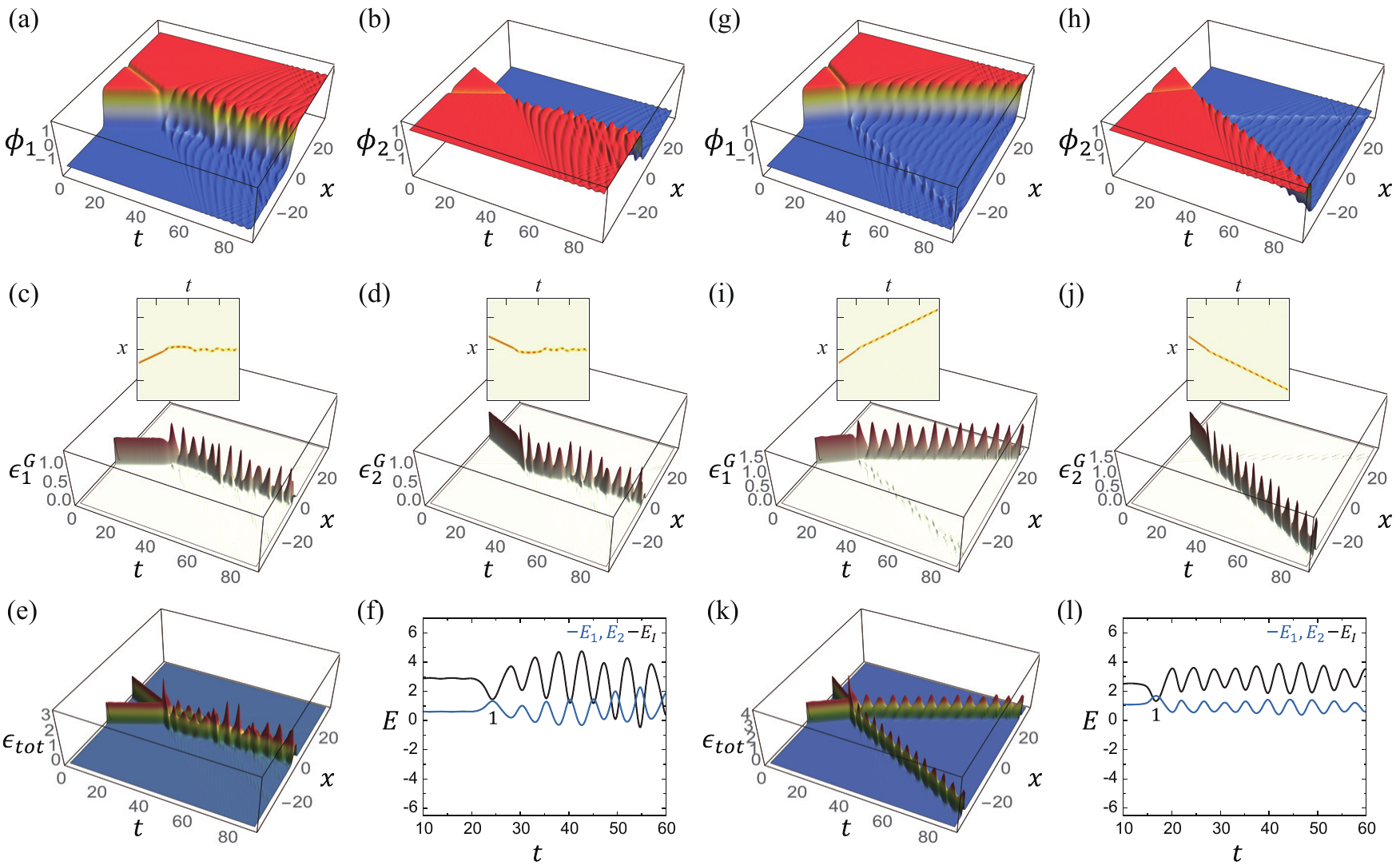}
\caption{
    \label{fig4:LC-LC} 
    \textbf{Collisions between two LC kinks.}
    \textbf{(a,b) [(g,h)]} The profiles of two fields,
    \textbf{(c,d) [(i,j)]} the gradient energy densities of two fields,
    \textbf{(e) [(k)]} the total energy density, and
    \textbf{(f) [(l)]} spatially-integrated energies $E_1$, $E_2$, and $E_I$ for $v_i=0.4$ ($v_i=0.6$).
    $\epsilon^{G}_{i}$ and $\epsilon_{tot}$ are the gradient energy density for the $i$th field and total energy density, respectively.
    In (c,d,i,j), each inset shows the top view of each gradient energy density.
    In (f,l), $E_i$, $E_I$, and the black number $n$ indicate the field energy, inter-field coupling energy, and moment of the $n$th collision.
}
\end{figure}
%

%%%%%%%%%%%%%%%%%%%%%%%%%%%%%%%%%%%%%%%%%%%%%%%%%%%%
%                      LC-LC
%%%%%%%%%%%%%%%%%%%%%%%%%%%%%%%%%%%%%%%%%%%%%%%%%%%%
\subsection{LC-LC and RC-RC kink collision\label{sec:LC-LC-RC-RC}}
Similar to the fact that the mirror symmetries in (\ref{eq:2.5:symmetry-1}) relate the RC and LC kinks, the collision between two RC kinks is also related to that between two LC kinks via the same mirror symmetries.
Therefore, we investigate only the LC-LC kink collision in this subsection.

\begin{figure}[t]
\centering
\includegraphics[width=.8\textwidth,origin=c]{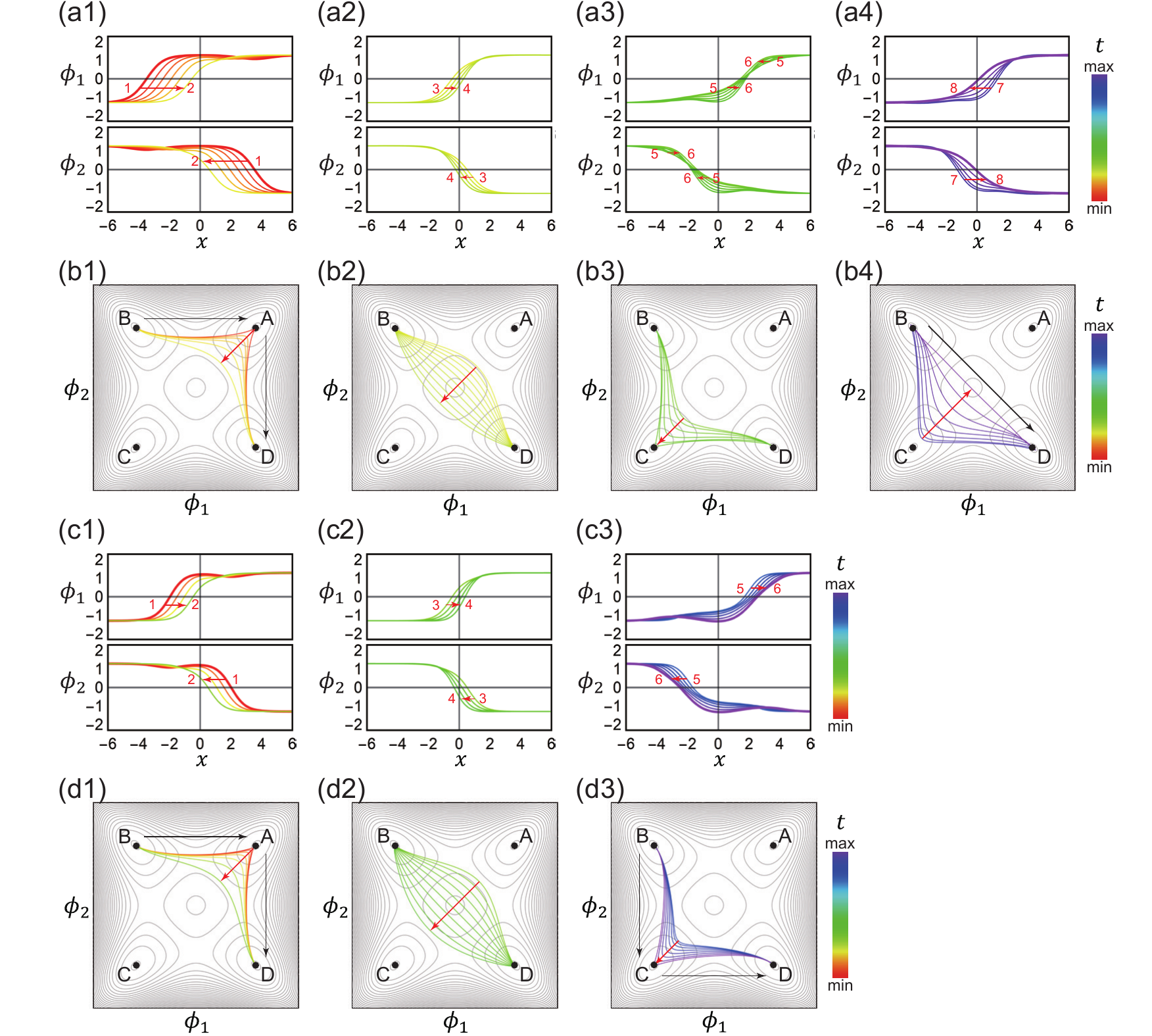}
\caption{
    \label{fig5:LC LC Contours} 
    \textbf{The collisions of two LC kinks in real and field spaces.}
    \textbf{(a) [(c)]} Time evolution of real-space profiles of the two fields for  $v_i=0.4$ ($v_i=0.6$).
    \textbf{(b) [(d)]} Time evolution of the orbits of the two fields in two-dimensional field space  for  $v_i=0.4$ ($v_i=0.6$). 
    In (b,d), the black arrows indicate the initial and final kinks.
    }
\end{figure}

%%%%%%%%%%%%%%%%%%%%%%%%%%%%%%%%%%%%%%%%%%%%%%%%%%%
%%%% 기본 설정과 충돌 결과에 대한 간단한 설명 %%%%%
%%%%%%%%%%%%%%%%%%%%%%%%%%%%%%%%%%%%%%%%%%%%%%%%%%%
As a representative example, we consider the $BA$ and $AD$ kinks collision.
We prepare an initial configuration such that $BA$ and $AD$ kink are located at the left- and right-hand sides, respectively, and move towards each other with the initial velocities $v_{i}$ (see figures~\ref{fig4:LC-LC} and \ref{fig5:LC LC Contours}):
\begin{align}
    & \phi_{1}(t,x)
    = \phi_{1}^{(B, A)}(\xi_{-}) + \phi_{1}^{(A, D)}(\xi_{+}) - \mu,
    \\
    & \phi_{2}(t,x)
    = \phi_{2}^{(B, A)}(\xi_{-}) + \phi_{2}^{(A, D)}(\xi_{+}) - \mu,
\end{align}
where $\xi_{\pm} =\gamma [ x \mp (x_0 - v_{i} t) ] $, $\gamma\equiv(1-v_{i}^2)^{-1/2}$, and  
$x_0$ is the initial position. 

\begin{figure}[t]
\centering
\includegraphics[width=.8\textwidth,origin=c]{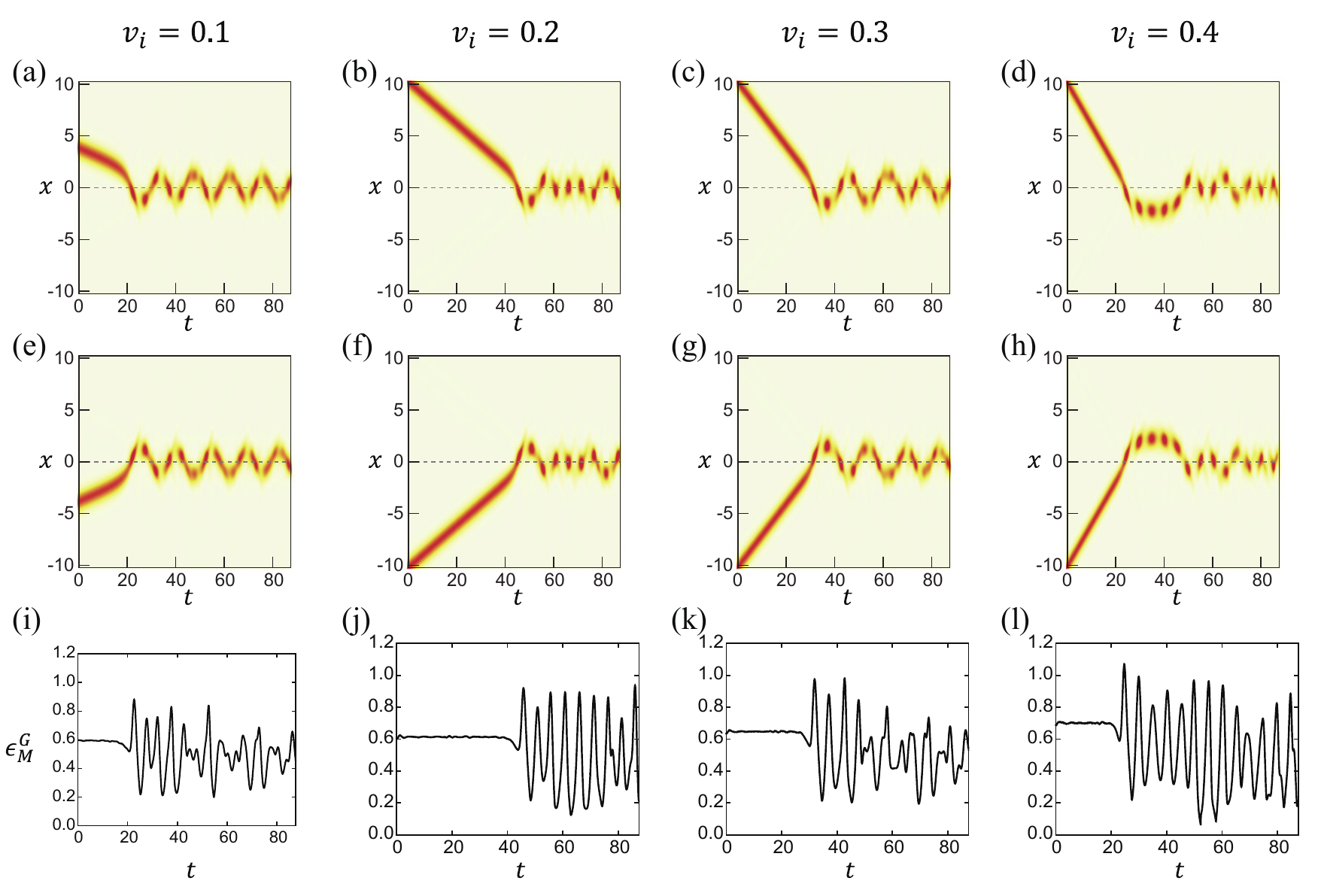}
\caption{
    \label{fig10:oscillation}
    \textbf{Bound mode and internal mode in the LC-LC kink collision below the critical velocities.}
    % \textbf{Bouncing motion of the bound of two kinks and internal mode oscillations of an NC kink during the LC-LC kink collisions.}
    \textbf{(a-h)}
    % Bouncing motion of the center of the two kinks.
    %Top views of the gradient energy densities $\epsilon_1^G$ and $\epsilon_2^G$ for  $\phi_1$ and $\phi_2$ are plotted in (a-d) and (e-h), respectively, which show the paths of the kinks clearly.
    Top views of the gradient energy densities $\epsilon_1^G$ and $\epsilon_2^G$ are plotted in (a-d) and (e-h), respectively, which show the paths of the kinks clearly.
    These paths show bouncing motions of two kinks.
    \textbf{(i-l)} 
    % The internal mode oscillations of NC kinks.
    % (i-l)
    The maximum value of the gradient energy density $\epsilon_M^G$ of a field for a given $t$ is plotted. $\epsilon_M^G$ of $\phi_1$ and $\phi_2$ are the same for the LC-LC kink collision. These plots show rather complex oscillations of internal modes of an NC kink due to the interference of the bouncing motions of two kinks after the collision.
    The initial velocities of the colliding two LC kinks are shown in the top. %, which are $v_i=0.1,0.2,0.3$, and $0.4$.
}
\end{figure}

The LC-LC kink collisions are separated by a critical velocity ($v_c=0.420$).
When $v_i > v_c$, two LC kinks pass through each other and become two RC kinks with internal modes after the collision [figures~\ref{fig4:LC-LC}(g-l) and \ref{fig5:LC LC Contours}(c,d)].
On the other hand, when $v_i < v_c$, two LC kinks are captured, forming a bound state (or an NC kink) with an internal mode [figures~\ref{fig4:LC-LC}(a-f) and \ref{fig5:LC LC Contours}(a,b)].
After the collision, small ripples are generated by the motion of the bound state as shown in figure~\ref{fig4:LC-LC}(a,b), which are continuum modes.
Because of this, the bound state becomes an NC kink with internal mode without bouncing,
while it is a stationary NC kink when time goes to infinity.
%%%%%%%%%%%%%%%%%%%%%%%%%%%%%%%%%%%%%%%%%%%%%
%%%% Fig6 Bound state and internal mode %%%%%
%%%%%%%%%%%%%%%%%%%%%%%%%%%%%%%%%%%%%%%%%%%%%
We distinguish between a bound state and an internal mode as shown in figure~\ref{fig10:oscillation}.
A bound state is composed of two primary kinks and the centers of the primary kinks show the bouncing motion [see the insets in figure~\ref{fig4:LC-LC}(c,d) and see also figure~\ref{fig10:oscillation}(a-h)].
On the other hand, the internal modes can be read from the oscillation of the amplitude of the gradient energy density in each field [see figures~\ref{fig4:LC-LC}(c,d) and \ref{fig10:oscillation}(i-l)].
% Note that, a bound state of two primar kinks in the same field, which will be seen in 
Note that, when two primary kinks in the same field form a bound state, we define such bound states a bion, similar to the single field model \cite{marjaneh2017multi}.
Furthermore, when the two induced lumps in the same field sufficiently overlap, they also form a small bion.

%%%%%%%%%%%%%%%%%%%%%%%%%%%%%%%%%%%%%%%%%%%%%%%%%%%%%%%%%%%%%%%%%%%%%
%% Figure 05 LC LC Contours
%%%%%%%%%%%%%%%%%%%%%%%%%%%%%%%%%%%%%%%%%%%%%%%%%%%%%%%%%%%%%%%%%%%%%

%% Figure 6 Oscillation and bouncing motion
%%%%%%%%%%%%%%%%%%%%%%%%%%%%%%%%%%%%%%%%%%%%%%%%%%%%%%%%%%%%%%%%%%%%%

%%%%%%%%%%%%%%%%%%%%%%%%%%%%%%%%%%%%%%%%%%%%%%%%%%%%
%%%% Orbit을 이용한 좀더 디테일한 process 설명 %%%%%
%%%%%%%%%%%%%%%%%%%%%%%%%%%%%%%%%%%%%%%%%%%%%%%%%%%%
To understand the chirality switching mechanism, we investigate field-space orbits of the colliding kinks with respect to time in figure~\ref{fig5:LC LC Contours}.
Since the colliding LC kinks are far away from each other initially, the field-space orbit of the two LC kinks form a line of $B \rightarrow A \rightarrow D$ in the field space [figure~\ref{fig5:LC LC Contours}(b1,d1)] and the forces and velocities are zero at the $B$, $A$, and $D$ groundstate; $\vec T (t,x)= \vec P(t,x)$ and $\vec V(t,x)=0$.
However, as the two initial LC kinks approach each other in real space [figure~\ref{fig5:LC LC Contours}(a1,a2,c1,c2)], the spatial region of the $A$ groundstate is reduced so that the points near $x=0$ experience a nonzero force because $|\vec T(t,x)|$ increase and exceed $|\vec P(t,x)|$ at those points [see detailed $\vec F$ and $\vec V$ in supplementary movie at $t \approx 15$ and $t \approx 12$ for $v_i=0.4$ and $v_i=0.6$, respectively]. 
If the entire region of the $A$ groundstate vanishes, then the field-space orbit starts to move toward the origin in field space as shown in figure~\ref{fig5:LC LC Contours}(b1,d1).
After that, $\vec{F}$ keeps driving the field-space orbit so that $\vec{V} $ keeps increasing until the orbit becomes the straight line $B\rightarrow D$ as shown in figure~\ref{fig5:LC LC Contours}(b2,d2).
With nonzero $\vec{V}$, the field-space orbits try to become $B \rightarrow C \rightarrow D$ for both cases of $v_i<v_c$ [figure~\ref{fig5:LC LC Contours}(b3)] and $v_i>v_c$ [figure~\ref{fig5:LC LC Contours}(d3)].
Even though the two field-space orbits of two different collisions are similar until $t \lesssim  t_c$, the collision results are different because $\vec F$ and $\vec V$ are different:
Both the velocity and force arrows direct towards the $C$ groundstate in the field space for $v_i > v_c$ [figure~\ref{fig11:Force and Velocity}(b)],
while they direct towards the $B \rightarrow D$ line in field space for $v_i < v_c$ [figure~\ref{fig11:Force and Velocity}(a)].
Hence, $\vec F$ pulls the field-space orbit toward the $C$ groundstate [figure~\ref{fig5:LC LC Contours}(d3)], while $\vec F$ pulls the field-space orbit back toward the straight line $B \rightarrow D $ [figure~\ref{fig5:LC LC Contours}(b4)].
As the result, the final states for $v_i>v_c$ and $v_i<v_c$ become two RC kinks and an NC kink, respectively.
We emphasize that the difference between the single-field $\phi^4$ model and double-field $\phi^4$ model.

Usually, in the single-field $\phi^4$ model, two colliding kinks are either reflected or captured \cite{campbell1983resonance,campbell1986solitary, goodman2005kink,goodman2007chaotic}.
On the contrary, in the coupled double-field $\phi^4$ model, the chirality of the kink changes after the collision, which is an interesting point in the multi-field systems.
%
%%%%%%%%%%%%%%%%%%%%%%%%%%%%%%%%%%%%%%%%%%%%%%%%%%%%
%                      RC-LC
%%%%%%%%%%%%%%%%%%%%%%%%%%%%%%%%%%%%%%%%%%%%%%%%%%%%
\subsection{RC-LC and LC-RC kink collision}
% Because an LC-RC collision can be obtained by applying a mirror symmetry operation in (\ref{eq:2.5:symmetry-1}) to an RC-LC collision, this subsection considers an RC-LC collision only. 
This subsection considers an RC-LC kink collision only, as an LC-RC  kink collision can be obtained by applying a mirror symmetry operation in (\ref{eq:2.5:symmetry-1}) to an RC-LC kink collision.

\begin{figure}[tbp]
\centering
\includegraphics[width=.8\textwidth,origin=c]{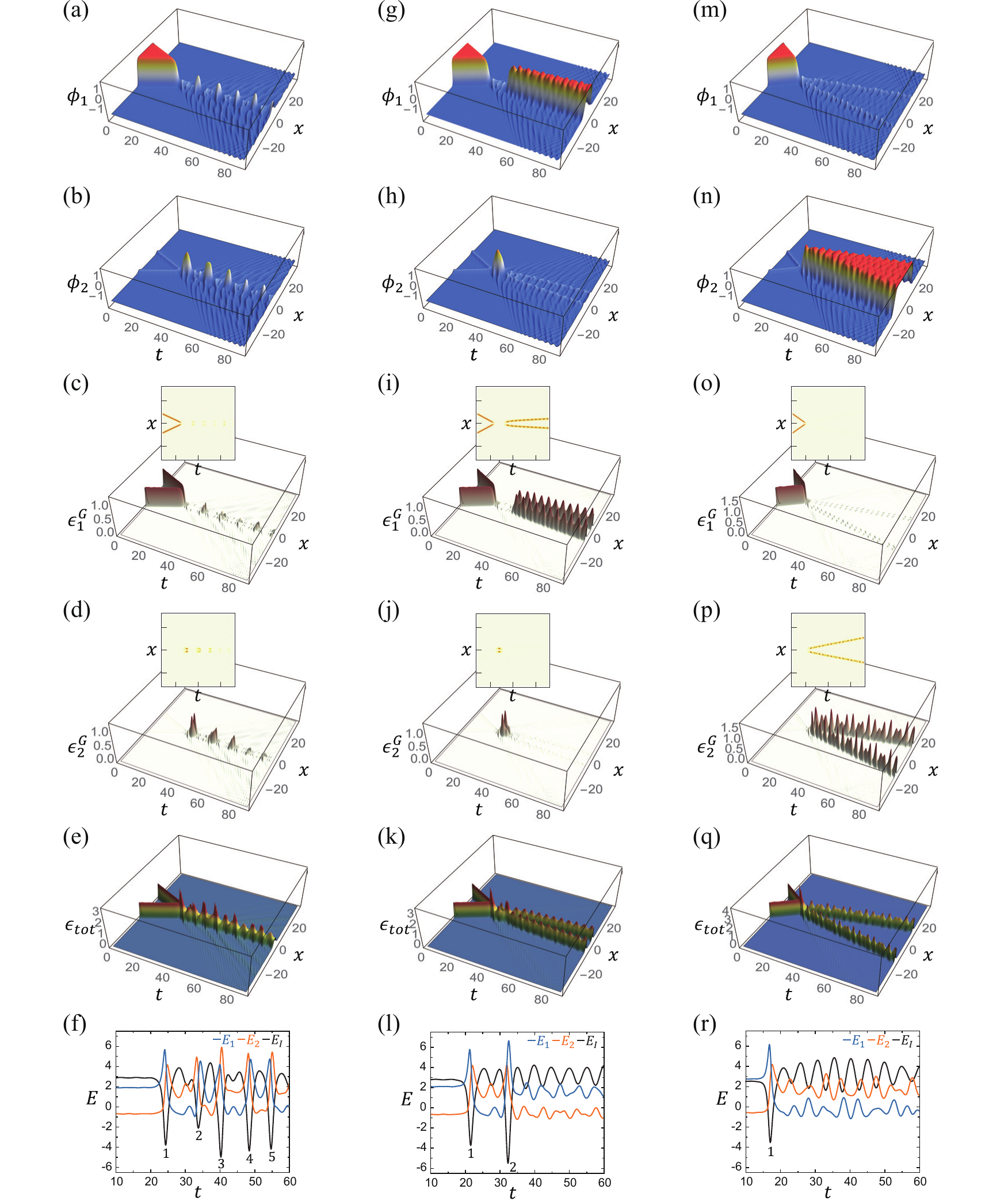}
\caption{
    \label{fig6:RC-LC} 
    \textbf{Collisions between RC-LC kinks.}
    The profiles, gradient energy densities of two fields, the total energy density, and spatially-integrated energies for \textbf{(a-f)} $v_i=0.4$, \textbf{(g-l)} $v_i=0.46$, and \textbf{(m-r)} $v_i=0.6$.
    The insets show the top views of the gradient energy densities.
 }
\end{figure}

\begin{figure}[tbp]
\centering
\includegraphics[width=.8\textwidth,origin=c]{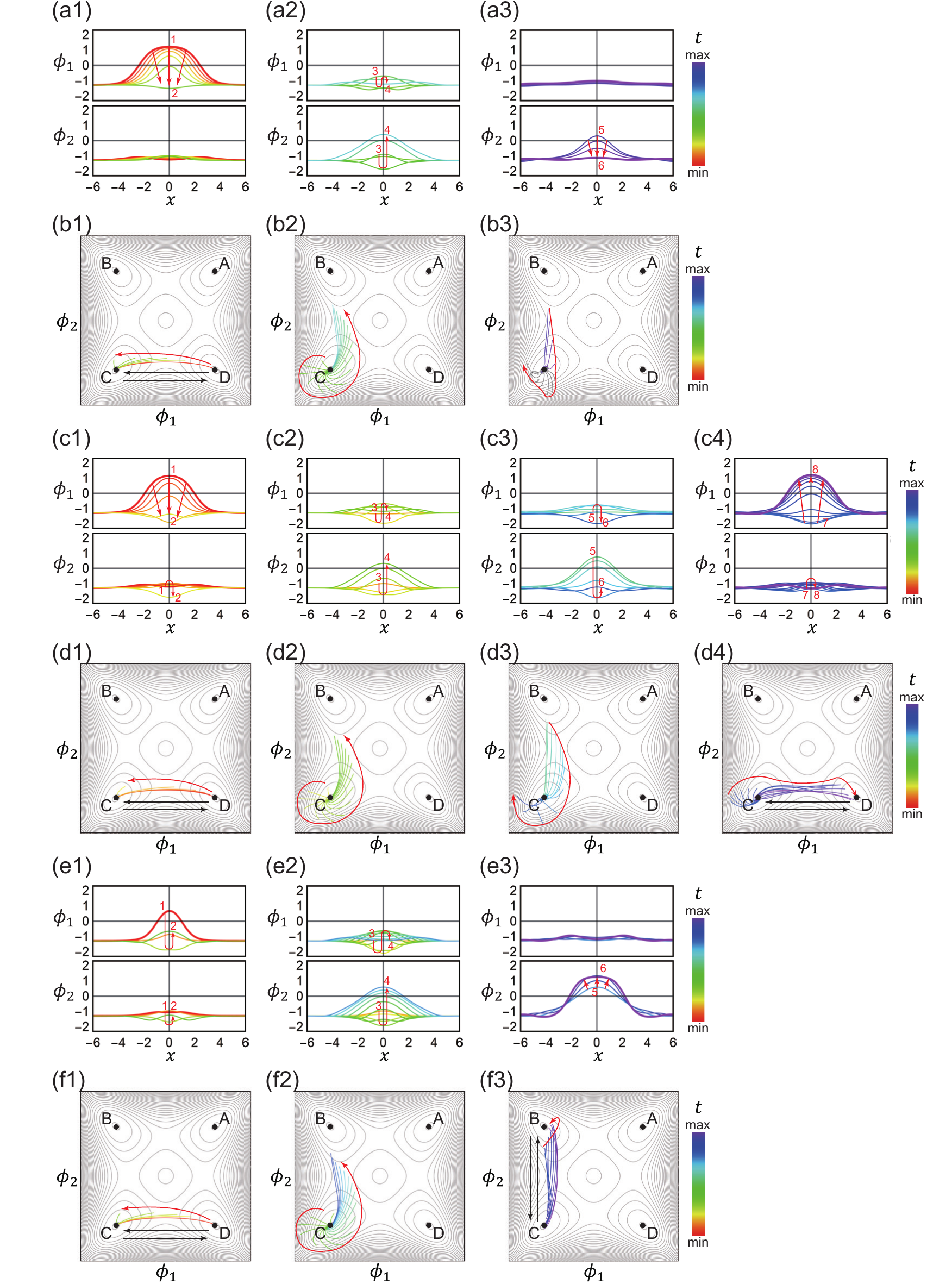}
\caption{
    \label{fig7:RC LC Contours} 
    \textbf{The collisions of RC and LC kinks in real and field spaces.}
    Real-space profiles of two fields for \textbf{(a)} $v_i=0.4$, \textbf{(c)} $v_i=0.46$, and \textbf{(e)} $v_i=0.6$.
    Time evolution of the orbits for the two fields in two-dimensional field space for \textbf{(b)} $v_i=0.4$, \textbf{(d)} $v_i=0.46$, and \textbf{(f)} $v_i=0.6$.
    }
\end{figure}

As a representative example, we consider the collision between $CD$ and $DC$ kinks moving towards each other with the initial velocities $v_i$ as shown in figures~\ref{fig6:RC-LC} and \ref{fig7:RC LC Contours}.
The initial configuration is given by
\begin{align}
    & \phi_{1}(t,x)
    = \phi_{1}^{(C,D)}( \xi_{-} ) + \phi_{1}^{(D,C)}( \xi_{+} ) - \mu,
    \\
    & \phi_{2}(t,x)
    = \phi_{2}^{(C,D)}( \xi_{-} ) + \phi_{2}^{(D,C)}( \xi_{+} ) + \mu,
\end{align}
where $\xi_{\pm} =\gamma [ x \mp (x_0 - v_{i} t) ] $ and $\gamma\equiv(1-v_{i}^2)^{-1/2}$. The initial primary RC and LC kinks reside in $\phi_1$ as shown in figure~\ref{fig7:RC LC Contours}(a1,c1,e1), and hence the initial configuration is similar to that of kink-antikink collisions in the single-field $\phi^4$ model. 
However, the results of the collision in this double-field $\phi^4$ model is different from those in the single-field $\phi^4$ model because of the interfield interaction. 
% , the specifics of which will be discussed later.
% 

The collision results are categorized into three cases depending on the initial velocity, as shown in figures~\ref{fig6:RC-LC} and \ref{fig7:RC LC Contours}.
First, when $v_i < v_c$ ($v_i \neq v_p$), the LC and RC kinks form a bion which has internal modes. Here, $v_c$ and $v_p$ are critical and particular velocities, respectively.
The second is when $v_i > v_c$, the primary LC and RC kinks in a field $\phi_i$ ($i=1,2$) become the primary RC and LC kinks in the other field $\phi_{3-i}$. 
Finally, when $v_i=v_p$ ($v_p < v_c$),
the initial primary LC and the RC kinks bounce off each other in the same initial field and escape to infinities due to a coherent energy transfer between two bions in different fields.
%

%%%%%%%%%%%%%%%%%%%%%%%%%%%%%%%%%%%%%%%%%%%%%%%%%%%%%%%%%%%%%%%%%%%%%
%% Figure 07 RC LC main
%%%%%%%%%%%%%%%%%%%%%%%%%%%%%%%%%%%%%%%%%%%%%%%%%%%%%%%%%%%%%%%%%%%%%
%
%%%%%%%%%%%%%%%%%%%%%%%%%%%%%%%%%%%%%%%%%%%%%%%%%%%%%%%%%%%%%%%%%%%%%
%% Figure 08 RC LC Contours
%%%%%%%%%%%%%%%%%%%%%%%%%%%%%%%%%%%%%%%%%%%%%%%%%%%%%%%%%%%%%%%%%%%%%
%
%%%%%%%%%%%%%%%%%%%%%%%%%%%%%%%%%%%%%%%%%%%%%%%%%%%%%%%%%%%%%%%%%%
%%%% Figure 9 Exchange energy of the two bions
%%%%%%%%%%%%%%%%%%%%%%%%%%%%%%%%%%%%%%%%%%%%%%%%%%%%%%%%%%%%%%%%%%
\begin{figure}[tbp]
\centering
\includegraphics[width=.8\textwidth,origin=c]{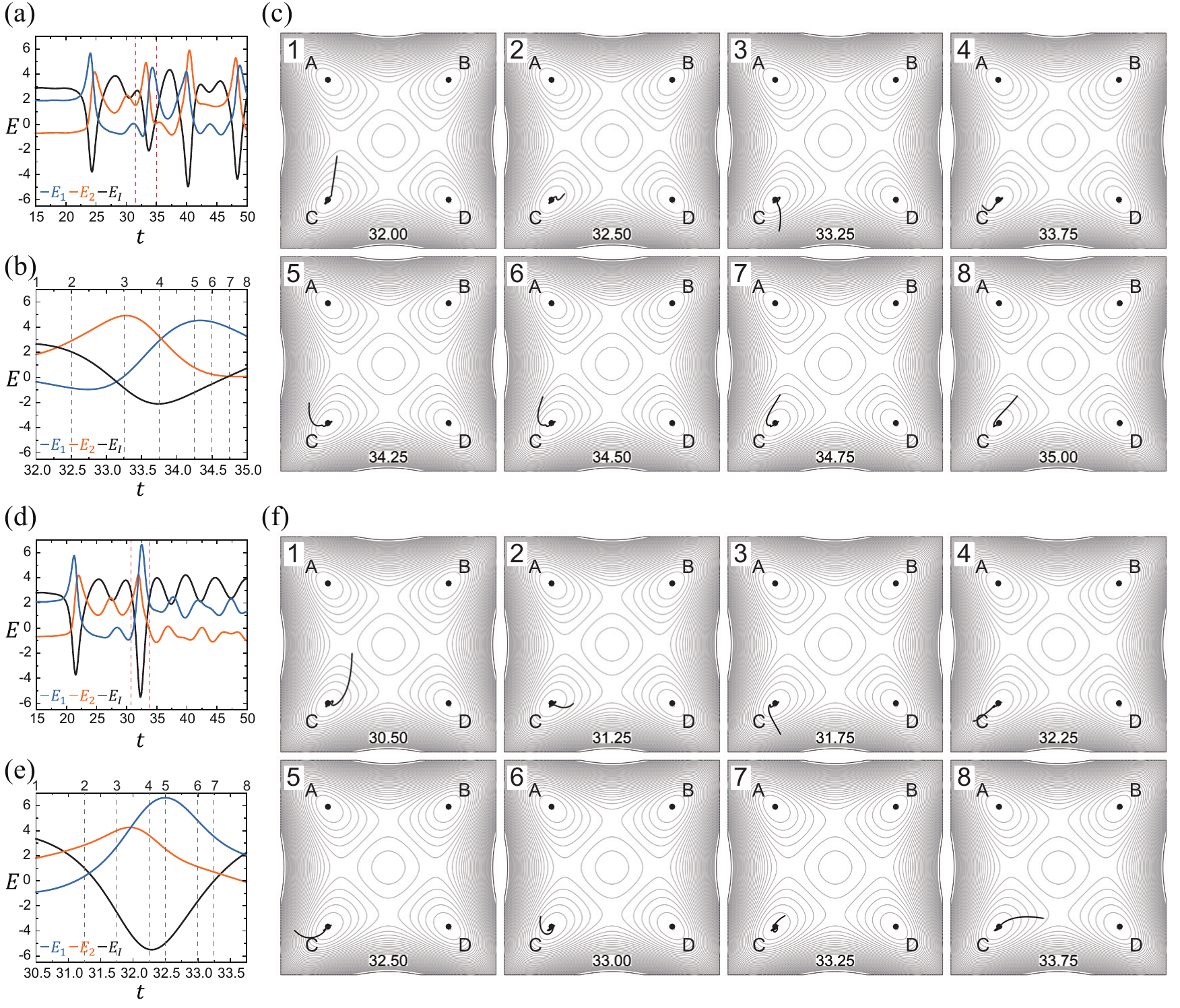}
\caption{
    \label{fig12:resonance}
    \textbf{Energy exchange of two bions in different fields during RC-LC kink collisions}
    % \textbf{Coherent vibration of two bions in different fields in an RC-LC kink collision}
    \textbf{(a)} \textbf{[(d)]} Spatially-integrated energy of each field $E_i[\phi_i]$, interfield potential energy $E_I[\vec \phi]$, and \textbf{(b)} \textbf{[(e)]} the closeup of the dashed red box for the RC-LC kink collision with the initial velocity $v_i=0.4$ ($v_i=0.46$). \textbf{(c)} \textbf{[(d)]} Several snapshots of the orbits during the collision interval in (b) [(e)].
    In (c,f), the numbers in the bottom indicate the snapshot time.
    (See the complete motions of orbits in supplementary movie online.)
    }
\end{figure}

%
%%%%%%%%%%%%%%%%%%%%%%%%%%%%%%%%%%%%%%%%%%%%%%%%%%%%%%%%
%%%%%%%%%%% 첫번째 충돌에 대한 일반적인 설명 %%%%%%%%%%%
%%%%%%%%%%%%%%%%%%%%%%%%%%%%%%%%%%%%%%%%%%%%%%%%%%%%%%%%
For the above three cases, collision processes are similar until the end of the first collision.
As the initial RC and LC kinks approach each other,
two primary kinks in $\phi_1$ and two induced lumps in $\phi_2$ form bions in their respective fields, and meet at $t=t_{c,1}$.
During the first collision, the bion in $\phi_1$ is pair-annihilated [figures~\ref{fig6:RC-LC}(a,g,m) and \ref{fig7:RC LC Contours}(a1,c1,e1)].
Then, the majority of the bion energy is transferred to the bion in $\phi_2$ via interfield coupling [see the peak labeled as $1$ in the $E_I$ plot of figure~\ref{fig6:RC-LC}(f,l,r)],
while a little energy still remains in the bion of $\phi_1$ [figure~\ref{fig7:RC LC Contours}(a2,c2,e2)].
The process so far is similar in all cases, but the subsequent result is different depending on $v_i$ because the detailed inelastic kink collision process is different.
%

%%%%%%%%%%%%%%%%%%%%%%%%%%%%%%%%%%%%%%%%%%%%%%%%%%%%%%%%%%%%
%%%%%%%%%%% Inelastic scattering에 관한 설명 %%%%%%%%%%%%%%%
%%%%%%%%%%%%%%%%%%%%%%%%%%%%%%%%%%%%%%%%%%%%%%%%%%%%%%%%%%%%
To understand the inelastic scattering process in the first collision, we consider the distributions of energies of colliding RC and LC kinks, which are stored in the bions, internal modes, and continuum modes in $\phi_i$ as well as in the interfield energy $E_I$.
During the collisions, we mainly consider the energies of the bions and the interfield energy because the energy of the bion in $\phi_i$ can be approximated to $E_i$ as the the energy of the internal and continuum modes are relatively small.
Meanwhile, if an elastic collision occurs, the relative speed of the scattered kinks are the same as the relative speed of the initial kinks.
Therefore, for an elastic collision to occur in the RC-LC kink collision case, the energy of the scattered bion must be the same as the energy of the colliding bion.
However, bions rarely experience elastic collision since the energy is exchanged between the two bions.
Despite the collisions being inelastic, the energy transfers can be distinguished as efficient or inefficient, depending on the ratio of energy transferred from one field to the other, and the amount of energy lost to the continuum and internal modes.
%

%%%%%%%%%%%%%%%%%%%%%%%%%%%%%%%%%%%%%%%%%%%%%%%%%%%%%%%%%%%%%%%%%%%%%%
%%%%%%%%%%% 첫번째 충돌에 대한 orbit 관점의 설명 (충돌 전) %%%%%%%%%%%
%%%%%%%%%%%%%%%%%%%%%%%%%%%%%%%%%%%%%%%%%%%%%%%%%%%%%%%%%%%%%%%%%%%%%%
To understand the efficient energy transfer between two bions, we discuss the first collision process focusing on an orbit in field space.
When RC and LC kinks come close to each other, one end of the initial orbit is lifted towards the origin, because of the bumps of $V_D$ between two groundstates [see red arrows the figure~\ref{fig7:RC LC Contours}(b1,d1,f1)].
A faster colliding velocity induces a stronger force in the direction to the lift of the orbit.
This is reflected in the real-space profile of induced lumps [see $\phi_2$ in figure~\ref{fig7:RC LC Contours}(a1,c1,e1)].
(We want to comment that this lifting process can be seen also in figure~\ref{fig7:RC LC Contours}(d3) which we will discuss in the second collision for the $v_i=v_p$ case.)
The lifting of the orbit prevents the orbit from moving straight to a groundstate, so that $\vec{T}$ generally induces a smooth winding [figures~\ref{fig7:RC LC Contours}(b2,d2,f2)].
This smooth winding has an important role in the efficient energy transfer between two bions in different fields which are as follows.
First, for efficient energy transfer, during the collision, $|\vec{\phi}|$ should be large and the orbit should sweep the region of $\theta \approx \frac{\pi}{4}$ considering the shape of the interfield potential $V_I$.
Here, $\theta \equiv \tan^{-1} (|\phi_2|/|\phi_1|)$.
Another reason is that if there is a fluctuation of the orbit length as shown in figure~\ref{fig7:RC LC Contours}(b3), there exists energy loss to internal and continuum modes.
Due to these reasons, the smooth winding process induces efficient energy transfer [see the figure~\ref{fig7:RC LC Contours}(b2,d2,f2)].
Moreover, such efficient energy transfer can be seen quantitatively in the $E_1$, $E_2$, and $E_I$ plots.
During the smooth winding at the first collision,
a strong peak in $E_I$ followed by the energy exchange between $E_1$ and $E_2$ is observed in figure~\ref{fig6:RC-LC}(f,l,r).
Note that, however, not all strong peaks in $E_I$ mean efficient energy transfer between two fields.
For example, during the LC-LC kink collision processes, efficient energy transfer between the two fields does not occur since $\theta$ hardly changes when $|\vec{\phi}|$ is large [see the orbits in figure~\ref{fig5:LC LC Contours}(b3,d3)]; 
quantitatively, $E_1=E_2$ as shown in figure~\ref{fig4:LC-LC}(f,l).
%

%%%%%%%%%%%%%%%%%%%%%%%%%%%%%%%%%%%%%%%%
%%%%%%%%%%% v_c 보다 큰 경우 %%%%%%%%%%%
%%%%%%%%%%%%%%%%%%%%%%%%%%%%%%%%%%%%%%%%
% We discussed the first collision so far.
We now investigate what happens after the first collision.
When $v_{i} > v_c$, due to the first smooth winding near $t=t_{c,1}$, the energy is transferred efficiently to $\phi_2$.
Although little energy is distributed to the internal and continuum modes, two kinks in the bion at $\phi_2$ are separated and escape to infinities since the initial kinks have enough energy.
As a result, the initial orbit $C \rightarrow D \rightarrow C$ becomes $C \rightarrow B \rightarrow C$  [see the orbit in figure~\ref{fig7:RC LC Contours}(f1-f3)].

%%%%%%%%%%%%%%%%%%%%%%%%%%%%%%%%%%%%
%%%%%%%%%% vc보다 작을 때 %%%%%%%%%%
%%%%%%%%%%%%%%%%%%%%%%%%%%%%%%%%%%%%
% Next, we consider collision processes satisfying $v_i < v_c$.
On the other hand, when $v_{i} < v_c$, 
the bion in $\phi_2$ cannot escape since energy is not enough, even though the majority of the energy is transferred from the bion in $\phi_1$ during the first collision.
After the escape is frustrated, the two primary kinks in $\phi_2$ are attracted to each other as shown in figure~\ref{fig7:RC LC Contours}(a3) [figure~\ref{fig7:RC LC Contours}(c3)], and the stretched green orbit toward the $B$ groundstate in figure~\ref{fig7:RC LC Contours}(b2) [figure~\ref{fig7:RC LC Contours}(d2)] is dragged back to the $C$ groundstate as shown in figures~\ref{fig7:RC LC Contours}(b3) [figures~\ref{fig7:RC LC Contours}(d3)], which leads to the second collision at $t=t_{c,2}\approx32.50$ ($t=t_{c,2}\approx31.25$) for $v_i=0.4$ ($v_i=0.46$).

In a usual case of $v_{i} < v_c$ ($v_i \neq v_p$), since the velocity of the primary kinks in $\phi_2$ is not enough, the lifting process is not sufficiently occurred to induce a smooth winding process [figure~\ref{fig7:RC LC Contours}(b3)].
Furthermore, the bion in $\phi_1$ does not help the orbit start winding the groundstate $C$ [figure~\ref{fig7:RC LC Contours}(b3)].
Therefore, energy is lost to continuum and internal modes during the fluctuated winding process, transferring energy inefficiently from the bion in $\phi_2$ to the bion in $\phi_1$ [see ripples in figure~\ref{fig6:RC-LC}(a,b)].
After that, energy is transferred from the bion in $\phi_1$ to the bion in $\phi_2$ in the same matter, and this process will be repeated.
This result shows a dissipative bound state [figures~\ref{fig6:RC-LC}(a-f)].
After enough time, the dissipative bound state becomes the groundstate $C$ due to dissipation via the continuum mode.
Note that the critical velocity of the double-field $\phi^4$ model ($v_c=0.56$) is greater than that of the single-field $\phi^4$ model ($v_c=0.26$ \cite{goodman2005kink}) due to the existence of added degrees of freedom to distribute energy.
% in the viewpoint of energy distribution.

%%%%%%%%%%%%%%%%%%%%%%%%%%%%%%%%%%%%%%%%%%%%%%%%%%%%%%%%%%%%%%%%%%%%%%%
%%%%%%%%%%%% vp 일 경우 (Fig12 : Energy exchange of two bions) %%%%%%%%%%%%
%%%%%%%%%%%%%%%%%%%%%%%%%%%%%%%%%%%%%%%%%%%%%%%%%%%%%%%%%%%%%%%%%%%%%%%
% Unlike the usual case of $v_i=v_p$ ,
Unlike the usual case of $v_i<v_c$, when $v_i=v_p$, at the second collision ($t_{c,2}\approx31.25$), the distance between two primary kinks in $\phi_1$ is sufficient such that the orbit starts to smoothly wind the groundstate $C$ [compare the motions of the orbit in figure~\ref{fig7:RC LC Contours}(b3) and (d3)], which leads to energy efficiently transferring from the bion in $\phi_2$ to the bion in $\phi_1$.
The reason why this smooth winding happens in the second collision is the coherent vibration, which will be discussed in the next paragraph.
As a result, the bion in $\phi_1$ acquires sufficient energy and escapes to infinities.
During $30.5 < t < 32.5 $, $E_2$ and $E_I$ are transferred to $E_1$ enabling a higher peak of $E_1$ than the first peak of $E_1$ at first collision as shown in figure~\ref{fig12:resonance}(d).
At $t=32.5$, $E_1$ is mostly distributed as the static energy $E_{S}$ in $\phi_1$ because the orbit is pulled to the left maximally and about to launch to the right in figure~\ref{fig12:resonance}(f).
On the other hand, the majority of the energy of $\phi_2$ is transferred to $\phi_1$
since it is the moment of pair-creation of two primary kinks in $\phi_2$.
After $t=32.5$, the orbit is cast towards the right with sufficient energy.
During this escape, $\phi_1$ loses energy to $V_I$ but the remnant energy of the bion in $\phi_2$ is transferred to $\phi_1$ by the bump of potential $V_D$ [see the figure~\ref{fig12:resonance}(e)].
Therefore, the pair-creation in $\phi_2$ is suppressed as shown in figure~\ref{fig7:RC LC Contours}(c4, d4).
As a result, the bion in $\phi_1$ receives sufficient energy and succeeds in escaping, which makes the final orbit $C \rightarrow D \rightarrow C$.
%

%%%%%%%%%%%%%%%%%%%%%%%%%%%%%%%%%%%%%%
%%%%%%% Two bounce collision %%%%%%%%%
%%%%%%%%%%%%%%%%%%%%%%%%%%%%%%%%%%%%%%
% 
The resonance energy exchange between a bion and internal mode has a significant role to explain the two bounce scattering via the reverse process in a single-field $\phi^4$ model \cite{campbell1983resonance,campbell1986solitary, goodman2005kink,goodman2007chaotic}.
However, the energy that can be stored in the internal mode is relatively small when we compare the internal mode with a bion not only in the single-field but also in the double-field $\phi^4$ model.
In the double-field $\phi^4$ model, each field can possess a bion which can exchange energy with their counterpart in the opposing field.
Hence, between the first and second collision, for the bion of $\phi_2$, the bion of $\phi_1$ stores and returns energy like an internal mode in the single-field model [see the vibration of the bion in $\phi_1$ in figure~\ref{fig7:RC LC Contours}(c2,c3)].
When the vibration phases of the two bions are appropriate, the bion in $\phi_1$ is prior to shrinking at the moment of pair-annihilation in $\phi_2$ ($t=31.25$) and the bion in $\phi_1$ preserves maximum energy at the moment of pair-creation in $\phi_2$ ($t=32.50$) as shown in figure~\ref{fig12:resonance}(f).
We call such vibrations between two bions in different fields as coherent vibrations, an example of which is shown in figure~\ref{fig7:RC LC Contours}(c2,c3,d2,d3).
However, there is one more condition for a two bounce scattering to occur.
While the energy loss from the internal mode oscillation can be ignored as perturbation, the energy loss from the bion vibration emits unignorable energy as a continuum mode [see the ripple in figure~\ref{fig6:RC-LC}(a,b)].
Therefore, to avoid losing too much energy, the bion in the $\phi_2$ cannot but vibrate only once in the exceptional scattering case ($v_i=v_p$).
On the other hand, the internal mode need only match the phase no matter how many times it oscillates for a two bounce scattering in the single-field $\phi^4$ model to occur.
As the result of the coherent vibration, the whole process in figure~\ref{fig7:RC LC Contours}(d1-d4) becomes near time-reversal.
~
%%%%%%%%%%%%%%%%%%%%%%%%%%%%%%%%%%%%%%%%%%%%%%%%%%%%%%%%%%%%%%%%%%%%%
%% Figure 10 RC NC main
%%%%%%%%%%%%%%%%%%%%%%%%%%%%%%%%%%%%%%%%%%%%%%%%%%%%%%%%%%%%%%%%%%%%%
\begin{figure}[tbp]
\centering
\includegraphics[width=.8\textwidth,origin=c]{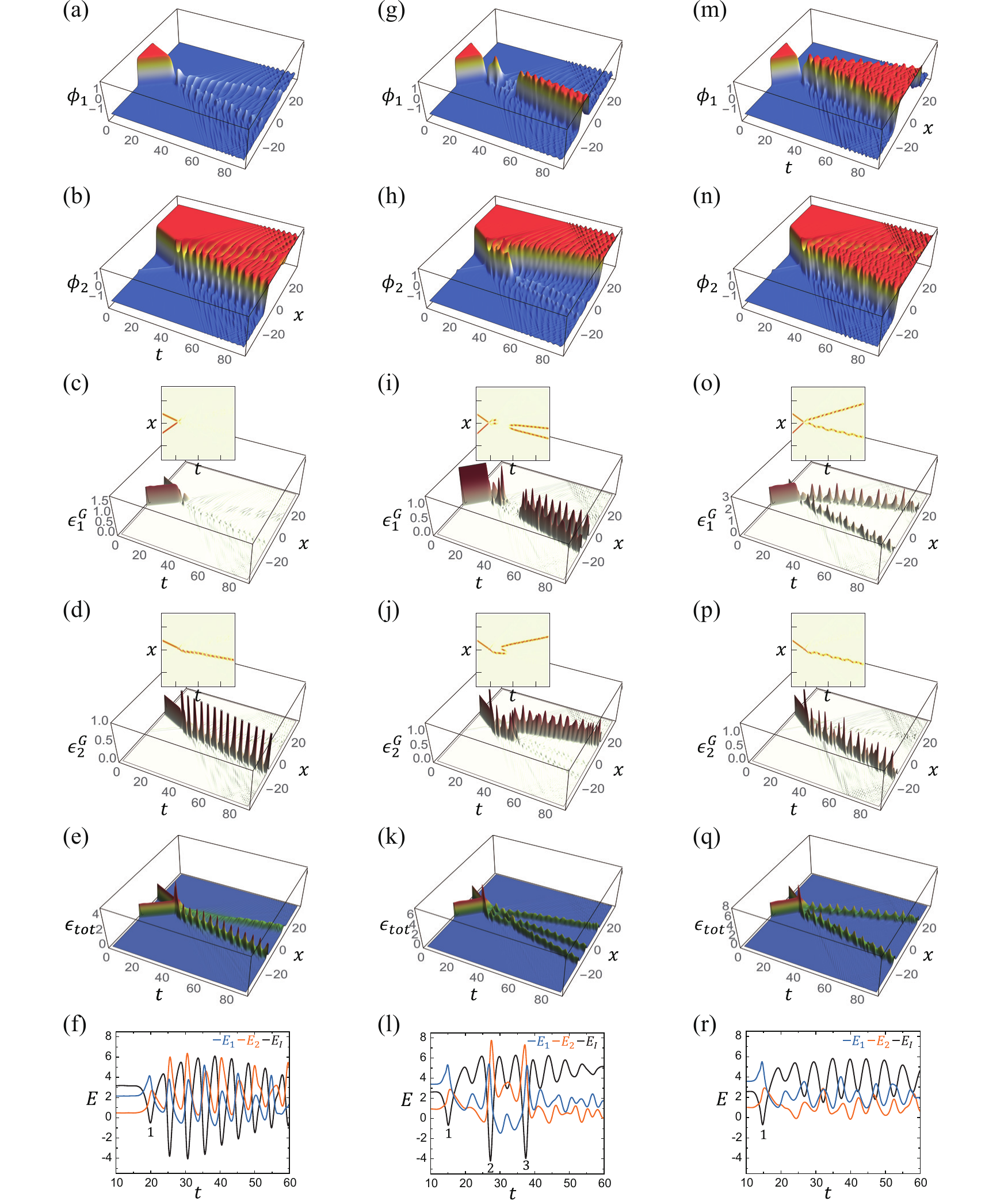}
\caption{
    \label{fig8:RC-NC} 
    \textbf{Profiles of collisions between RC and NC kinks.}
    The profiles, gradient energy densities of two fields, the total energy density, and spatially-integrated energies for \textbf{(a-f)} $v_i=0.6$, \textbf{(g-l)} $v_i=0.785$, and \textbf{(m-r)} $v_i=0.8$.
    The insets show the top views of the gradient energy densities.
}
\end{figure}
%
%%%%%%%%%%%%%%%%%%%%%%%%%%%%%%%%%%%%%%%%%%%%%%%%%%%%%%%%%%%%%%%%%%%%%
%% Figure 11 RC NC Contours
%%%%%%%%%%%%%%%%%%%%%%%%%%%%%%%%%%%%%%%%%%%%%%%%%%%%%%%%%%%%%%%%%%%%%
\begin{figure}[tbp]
\centering
\includegraphics[width=.8\textwidth,origin=c]{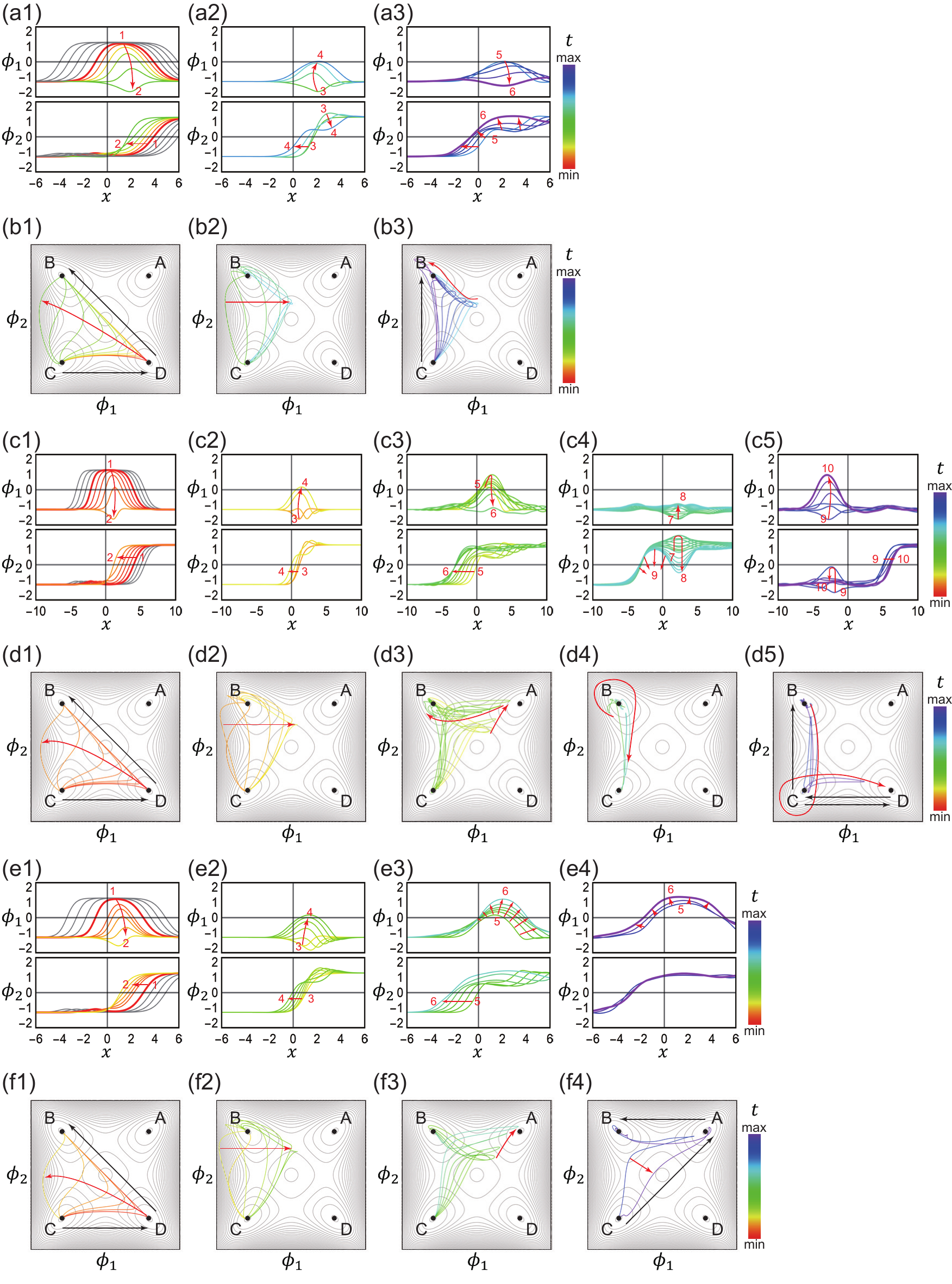}
\caption{
    \label{fig9:RC NC Contours} 
    \textbf{The collisions of RC and NC kinks in real and field spaces.}
    Real-space profiles of two fields for \textbf{(a)} $v_i=0.6$, \textbf{(c)} $v_i=0.785$, and \textbf{(e)} $v_i=0.8$.
    Time evolution of orbits for the two fields in two-dimensional field space for \textbf{(b)} $v_i=0.6$, \textbf{(d)} $v_i=0.785$, and \textbf{(f)} $v_i=0.8$.
    }
\end{figure}
%

%%%%%%%%%%%%%%%%%%%%%%%%%%%%%%%%%%%%%%%%%%%%%%%%%%%%%%%%%%%%%%%%%%%%%
%% Figure 12
%%%%%%%%%%%%%%%%%%%%%%%%%%%%%%%%%%%%%%%%%%%%%%%%%%%%%%%%%%%%%%%%%%%%%
\begin{figure}[tbp]
\centering
\includegraphics[width=.8\textwidth,origin=c]{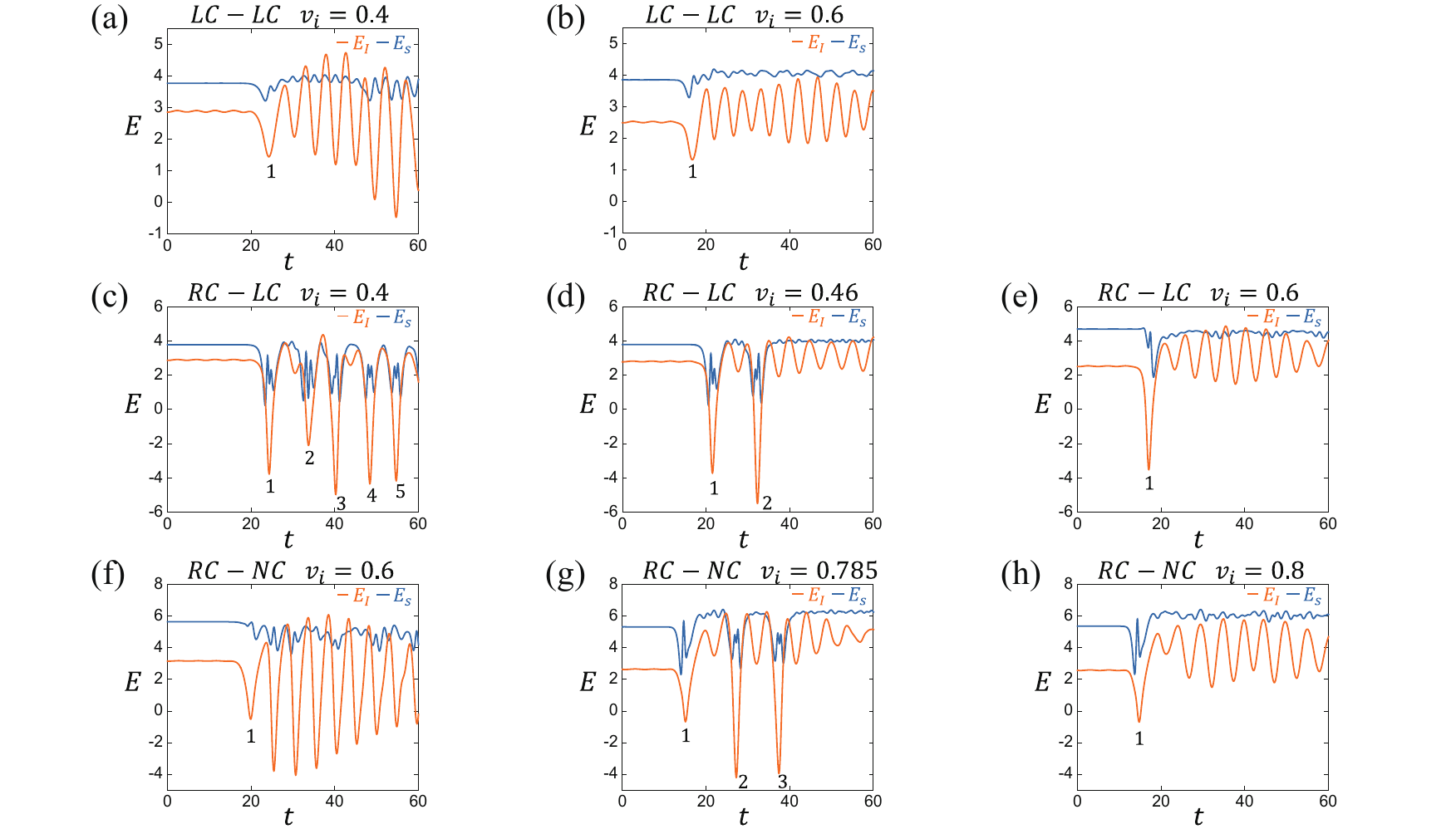}
\caption{
    \label{fig17:EI-ES} 
    \textbf{Time evolution of static energies under various kink collisions.}
    \textbf{(a,b)} LC-LC kink collisions.
    \textbf{(c-e)} RC-LC kink collisions.
    \textbf{(f-h)} RC-NC kink collisions.
    Here, $E_s$, $E_I$, and the black number $n$ indicate the static energy, inter-field coupling energy, and moment of the $n$th collision.
    }
\end{figure}
%
%%%%%%%%%%%%%%%%%%%%%%%%%%%%%%%%%%%%%%%%%%%%%%%%%%%%
%                      RC-NC
%%%%%%%%%%%%%%%%%%%%%%%%%%%%%%%%%%%%%%%%%%%%%%%%%%%%
\subsection{RC-NC kink collision}
This subsection considers only an RC-NC collision, because the other LC-NC, NC-RC, and NC-LC collisions can be obtained from an RC-NC collision by applying a mirror symmetry operation in (\ref{eq:2.5:symmetry-1}).

As a representative example, we study the collision between $CD$ and $DB$ kinks moving towards each other with the initial velocities $v_i^{RC}$ and $v_i^{NC}$, respectively, as shown in figures~\ref{fig8:RC-NC} and \ref{fig9:RC NC Contours}. 
The initial configuration is given by 
\begin{align}
    & \phi_{1}(t,x)
    = \phi_{1}^{(C,D)}( \xi^{RC} ) + \phi_{1}^{(D,B)}( \xi^{NC} ) - \mu,
    \\
    & \phi_{2}(t,x)
    = \phi_{2}^{(C,D)}( \xi^{RC} ) + \phi_{2}^{(D,B)}( \xi^{NC} ) + \mu,
\end{align}
where $\xi^{RC} =\gamma^{RC} [ x + (x_0 - v_i^{RC} t) ] $ and
$\xi^{NC} =\gamma^{NC} [ x - (x_0 - v_i^{NC} t) ]$ with $\gamma^{a} \equiv (1-(v_i^{a})^2)^{-1/2}$ ($a=RC, NC$).
The initial velocities ($v_i^{RC}$ and $v_i^{NC}$) are adjusted so that the total linear momentum is zero at the center-of-mass frame.
\begin{align}
    p^{RC} + p ^{NC} = \gamma^{RC} v_i^{RC} M_{0}^{RC} - \gamma^{NC} v_i^{NC} M_{0}^{NC}=0,
\end{align}
where $M_{0}^{RC}=1.88$ and $M_{0}^{NC}=3.14$ are the rest masses of the RC and NC kinks, respectively.

Depending on the initial velocities, the collision results are categorized into three cases.
Firstly, there exists a critical velocity ($v_c^{RC}=0.791$) that divides the collision regions into two parts.
When $v_i^{RC} > v_c^{RC}$, the initial RC and NC kinks become NC and RC kinks after the collision.
When $v_i^{RC} < v_c^{RC}$ but $v_i^{RC} \neq v_p^{RC}$, the initial RC and NC kinks become an LC kink after the collision.
When $v_i^{RC}$ is equal to a particular velocity, $v_i^{RC}= v_p^{RC}<v_c^{RC}$, the collision result is rather complex; initial RC and NC kinks become RC, LC, and LC kinks after the collision.
We study the details of each case. 

First, we study the collision process where $v_i^{RC} < v_c^{RC}$ but $v_i^{RC} \neq v_p^{RC}$.
Figures~\ref{fig8:RC-NC}(a-f) and \ref{fig9:RC NC Contours}(a,b) show the numerically calculated collision result for $(v^{RC}_i, v^{NC}_i) =  (0.60, 0.41)$. 
The initial RC and NC kinks approach each other and meet at the collision time of $t_{c,1}$ [figure~\ref{fig8:RC-NC}(a-f)].
When two kinks meet, the primary RC and NC kinks located in $\phi_1$ is pair-annihilated [figure~\ref{fig9:RC NC Contours}(a1)], and the initial orbit $C \rightarrow D \rightarrow B$ in field space is pulled toward the line of $C \rightarrow B$ [figures~\ref{fig9:RC NC Contours}(b1) and \ref{fig11:Force and Velocity}(g)].
Then, the remaining kinetic energy transforms to the potential energy and then the potential energy turns back to the kinetic energy [figure~\ref{fig9:RC NC Contours}(a1,a2)]. In this process, the moving direction of the orbit is reversed [compare the red arrows in figure~\ref{fig9:RC NC Contours}(b1) and \ref{fig9:RC NC Contours}(b2)].
%
%%%%%%%%%  add start %%%%%%%%%%%%%
Although the above collision and pair-annihilation process of the two primary kinks in $\phi_1$ is similar to those in the RC-LC kink collision cases, the primary kinks are not transferred to the other field in figure~\ref{fig9:RC NC Contours}(a2).
The existence of a primary kink in $\phi_2$ prevents the field-space orbit from winding around any groundstate in this collision [see orbit in figure~\ref{fig9:RC NC Contours}(b1)], while the field-space orbits always wind the groundstate $C$ in RC-LC kink collisions after the pair-annihilation of two primary kinks in $\phi_1$ [see the figure~\ref{fig7:RC LC Contours}(b2,d2,f2)].
This phenomenon can be seen irrespective of $v_i$ [see also the orbits in figure~\ref{fig9:RC NC Contours}(d1,f1)].
%%%%%%%%%  add end  %%%%%%%%%%%%%
%
The kinetic energy induces the pair-creation of two primary kinks in the same field, $\phi_1$ [figure~\ref{fig9:RC NC Contours}(a2)], and the orbit $C \rightarrow B$ is dragged to the groundstate $A$ [figure~\ref{fig9:RC NC Contours}(b2)].
However, because the kinetic energy is not enough, the two primary kinks in $\phi_1$ becomes a groundstate [figure~\ref{fig9:RC NC Contours}(a3)], whereas the primary kink in $\phi_2$ keeps its motion and becomes an LC kink [figure~\ref{fig9:RC NC Contours}(a3)].
At $t=t_{c,1}$, most of the energy in $\phi_1$ is transferred to $\phi_2$ via the interfield coupling and generates the oscillating internal mode in $\phi_2$ [figure~\ref{fig8:RC-NC}(d)].
In field space, the initial orbit $C \rightarrow D \rightarrow B$ [figure~\ref{fig9:RC NC Contours}(b1)] becomes the orbit $C \rightarrow B$ [figure~\ref{fig9:RC NC Contours}(b3)].
Therefore, the initial RC and NC kinks become an LC kink after the collision.

Second, we study the case where $v_i^{RC} > v_c^{RC}$.
Figures~\ref{fig8:RC-NC}(m-r) and \ref{fig9:RC NC Contours}(e,f) show the numerically calculated results for $(v^{RC}_i, v^{NC}_i) =  (0.785, 0.605)$. 
As shown in figures~\ref{fig8:RC-NC}(m-r) and \ref{fig9:RC NC Contours}(e1,e2), the collision process until the first collision ($t=t_{c,1}$) is similar to the above case [compare figure~\ref{fig9:RC NC Contours}(a1,a2) and figure~\ref{fig9:RC NC Contours}(e1,e2)]. 
However, because the kinetic energy is enough to overcome the attractive interaction,
two pair-created primary kinks in $\phi_1$ escape each other [figure~\ref{fig9:RC NC Contours}(e3)]. 
Moreover, the primary kink in $\phi_2$ helps the primary kink at the left side in $\phi_1$ to escape easily by an attractive interaction between them [figure~\ref{fig9:RC NC Contours}(e3)].
(The attractive interaction between two primary kinks was discussed in the previous LC-LC kink collision.)
Finally, the two primary kinks at the left side in $\phi_1$ and $\phi_2$ are combined and become an NC kink [figure~\ref{fig9:RC NC Contours}(e4)].
In field space, the initial orbit $C \rightarrow D \rightarrow B$ is dragged enough toward the groundstate $A$ [figure~\ref{fig11:Force and Velocity}(i)] and becomes the final orbit $C \rightarrow A \rightarrow B$, which results in NC and RC kinks [figure~\ref{fig9:RC NC Contours}(f1-f4)].
Therefore, the initial RC and NC kinks become NC and RC kinks.
%
%%%%%%%%%%%%%%%%%%%%%%%%%%%%%%%%%%%%%%%%%%%%%%%%%%%%%%%%%%%%%%%%%%%%%
%% Figure 16 NC NC add
%%%%%%%%%%%%%%%%%%%%%%%%%%%%%%%%%%%%%%%%%%%%%%%%%%%%%%%%%%%%%%%%%%%%%
\begin{figure}[tbp]
\centering
\includegraphics[width=.8\textwidth,origin=c]{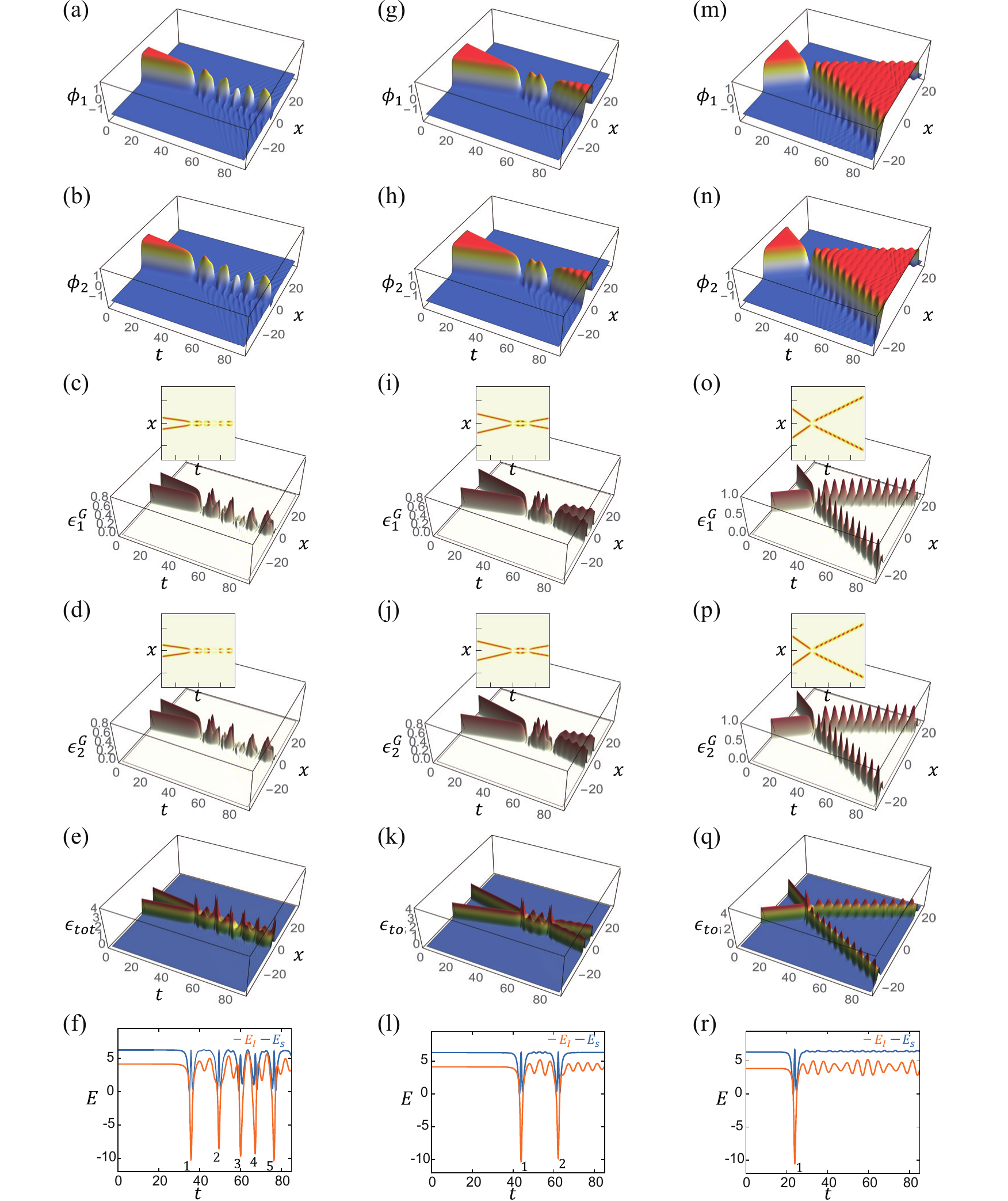}
\caption{
    \label{fig16:NC-NC} 
    \textbf{Collisions between NC-NC kinks.}
    Initial prepared $CA$ and $AC$ kinks collide with each other.
    The profiles, gradient energy densities of two fields, the total energy density, and spatially-integrated energies for \textbf{(a-f)} $v_i=0.12$, \textbf{(g-l)} $v_i=0.2$, and \textbf{(m-r)} $v_i=0.4$ are plotted.
    In (a-f), two NC kinks are captured and form a groundstate.
    In (g-l), two NC kinks experience $2$-bounce resonance and escape to infinity.
    In (m-r), two NC kinks reflect each other and escape to infinity.
    }
\end{figure}

Finally, we study the case where $v_i^{RC} = v_p^{RC} < v_c^{RC}$.
Figures~\ref{fig8:RC-NC}(g-l) and \ref{fig9:RC NC Contours}(c,d) show the numerically calculated results for $(v^{RC}_i, v^{NC}_i) = (0.800, 0.624)$. 
The collision process until the first collision ($t=t_{c,1}$) is similar to the previous case of $v_i^{RC} > v_c^{RC}$ [compare figure~\ref{fig9:RC NC Contours}(a1-a3) and figure~\ref{fig9:RC NC Contours}(e1-e3)].
Unlike the $v_i^{RC} = 0.6$ case, the two primary kinks in $\phi_1$ undergoes a longer process to collide a second time, resulting in the primary kink in $\phi_2$ to move further away [compare figure~\ref{fig9:RC NC Contours}(c2,c3) and figure~\ref{fig9:RC NC Contours}(a2,a3)].
The primary kinks in $\phi_1$ are unable to escape, unlike their faster counterparts [compare figure~\ref{fig9:RC NC Contours}(c3) and figure~\ref{fig9:RC NC Contours}(e3)].
After that, the two pair-created primary kinks in $\phi_1$ are captured and pair-annihilated again, as shown in the process from 5 to 6 in $\phi_1$ of figure~\ref{fig9:RC NC Contours}(c3).
This is the second collision ($t=t_{c,2}$).
Contrast to the first collision, at the moment of this pair-annihilation ($t=t_{c,2}$), the total energy is transferred to $\phi_2$, and two primary kinks are pair-created in $\phi_2$, since the primary kink in $\phi_2$ is far from the point of pair-creation, as shown in the process from 7 to 8 in $\phi_2$ of figure~\ref{fig9:RC NC Contours}(c4).
The left one between the two primary kinks in $\phi_2$ is pair-annihilated with the initially existing primary kink in $\phi_2$ [see the process of 9 in figure~\ref{fig9:RC NC Contours}(c4)], which leads to the third collision ($t=t_{c,3}$).
As a result of the third collision, the energy is transferred to $\phi_1$ again, and two primary kinks are generated in $\phi_1$, as shown in the process from 9 to 10 in figure~\ref{fig9:RC NC Contours}(c5).
(Note that the orbit winding a groundstate process at these second and third collisions can also be seen in the previous RC-LC kink collision.)
Finally, one gets the three separated kinks (RC, LC, and LC kinks).
In the field space, the initial orbit $C \rightarrow D \rightarrow B$ becomes the final orbit $C \rightarrow D \rightarrow C \rightarrow B$ (see the complete motion of orbits in supplementary movie online).
Therefore, the initial RC and NC kinks become RC, LC, and LC kinks.
Note that figure~\ref{fig8:RC-NC}(l) shows three strong peaks, which indicates the number of energy transfer between fields, which supports the above explanation.

\subsection{NC-NC kink collision}
For an NC-NC kink collision, $\phi_1(t,x) = \pm \phi_2 (t,x)$.
Thus, the field equation for each field $\phi_i$ reduces the field equation for the single-field $\phi^4$ model in (\ref{eq:2.3:single field equation}) except for the coupling constants:
$
    \partial_t^2 \phi-\partial_x^2 \phi - m^2 \phi + (\lambda+\alpha) \phi^3  = 0.
$
Then, the results of the NC-NC kink collision are the same with the reported results of the single-field $\phi^4$ model \cite{campbell1983resonance, campbell1986solitary, goodman2005kink, goodman2007chaotic}. 
First, the collision result depends on the initial velocity $v_i$ of the colliding NC kinks.
Next, there exist a critical velocity $v_c \approx 0.194$ which separates the collision regions.
When $v_i > v_c $, two incident NC kinks always escape to infinity after a collision [figure~\ref{fig16:NC-NC}(a-f)].
When $v_i < v_c$, two NC kinks are usually trapped and pair-annihilated [figure~\ref{fig16:NC-NC}(m-r)].
For $v_i=v_p$, we observe the so-called $n$-bounce resonance [see two-bounce scattering in figure~\ref{fig16:NC-NC}(g-l)].
Since the collision results are already reported in the single-field $\phi^4$ model \cite{ campbell1983resonance, campbell1986solitary, goodman2005kink, goodman2007chaotic}, we will not discuss the NC-NC kink collisions any further in this work.

\section{Summary and conclusion}
We investigate the collisions among chiral and nonchiral kinks in the coupled double-field $\phi_4$ model, and show that the kink collision satisfy the $Z_4$ abelian group operation.

Unlike the single-scalar-field $\phi_4$ model, our model has four degenerate vacua and twelve kinks interpolating two vacua.
We show that the twelve kinks are classified as right-chiral (RC), left-chiral (LC), and nonchiral (NC) kinks depending on the topological chiral charge and the mirror symmetry between $\phi_1$ and $\phi_2$;
all NC kinks have the topological chiral charge of $Q^{NC} = \pm 2$ (mod 4) and the mirror symmetry, $\phi_1 \leftrightarrow \phi_2$, up to the $Z_4$ field rotation symmetry.
On the other hand, RC and LC kinks break such mirror symmetry, and hence they have opposite topological chiral charges; $ Q^{RC} = - Q^{LC} = 1$ (mod 4).
Thus, we demonstrate that chiral and nonchiral kinks, including vacua, carry the quaternary topological chiral charges. %, which is the basis of the $Z_4$ collision.
Moreover, the model possesses $Z_4$ rotation symmetry ($\phi_1 \rightarrow - \phi_2$, $\phi_2 \rightarrow \phi_1$), which is one of the essential ingredients for the $Z_4$ collision among chiral and nonchiral kinks.
We find the solutions of such kinks and internal modes trapped in each kink, which contribute to the kink collisions being inelastic.

In principle, all collisions that preserve total topological chiral charge can be considered.
However, because not all collisions happen, we study the dynamical process and mechanism of all possible collisions between two kinks by solving the field equation numerically.
We also investigate energy densities, energy transfer between the two fields, and the time-evolution of orbits in two-dimensional field space $(\phi_1, \phi_2)$, which explain the collision mechanism and chirality change of kinks.
In particular, we find that kink collisions follow $Z_4$ abelian group operation table in figure~\ref{fig2:kinks} dynamically.
We show that the force and velocity vector fields defined in the field space and the energy exchange between different fields play an essential role in determining the fate of the kink collision.
The collision results are categorized into three cases depending on the initial velocity, $v_i$, summarized as table~\ref{table1:collision-table}.

When $v_i < v_c $ ($v_i \neq v_{p}$), we observe the formation of vibrating bound state of the two kinks,
which is either a bion having zero topological chiral charge or a kink having a nonzero topological chiral charge depending on the initial configuration.
If the final state is a bion, it decays into a groundstate emitting bosons.
On the other hand, if the final state is a kink, it has an oscillating internal mode.

When $v_i > v_c$, inelastic scatterings occur because the energy in a field can be distributed into the other field, internal modes, and continuum modes during the collision process.
Since the kinetic energy of two colliding kinks is large enough, they overcome the binding energy and escape to infinities.
If the topological charges of initial kinks are the same, then the scattered kinks have the opposite sign of topological charges of initial kinks.
On the other hand, when the chiralities of initial kinks are different,
the chirality set of the kinks does not change before and after the collision.
Interestingly, in RC-LC kink collisions, two initial primary kinks in one field are shifted to the other field, which is explained by the energy transfer between two fields via interfield interaction.

When $v_i = v_p$ ($v_i<v_c$), we observe an exceptional scattering, which could be characterized by multiple times of the energy exchange between two fields and kinks escaping from the collision point at the final stage.
For RC and LC kink collision, at the first collision, the colliding primary RC and LC kinks in a $\phi_i$ field form a bion but the majority of the energy is transferred to another $\phi_{3-i}$ field creating a bion (composed of two primary kinks) via the interfield coupling.
At the second collision, the two primary kinks of the bion in $\phi_{3-i}$ collide again and hence the bion shrinks while coherently transferring most of the energy to the remnant bion in $\phi_i$, generating two escaping primary kinks.
Thus, the coherent vibration of two bions results in a two-bounce kink scattering.
%
%%%%%%%%%%%% RC-NC %%%%%%%%%%%%%%
%
On the other hand, for an RC-NC kink collision, there exist one self-energy exchange in a field and two energy transfers between the fields during the collision process.
These energy exchange phenomena between fields can be applied to a wide class of multifield nonlinear systems.

We would like to emphasize the importance of the multifield theory and its kink dynamics because such topological objects and their dynamics appear in quantum and classical field systems, supersymmetric systems, high energy physics, cosmological systems, condensed matter systems, etc.
In this viewpoint, our work has made improvements in the area of multifield theory by investigating the nontrivial double-field model and its kink dynamics involving chirality and $Z_4$ abelian group operation.
Therefore, we expect that our work will be useful to understand the dynamics of topological objects in the multifield nonlinear system.
We also believe that the concept of chirality and multi-digit topological charge in this work have potential applications to topological information science.

Finally, we would like to mention several subjects for possible future study.
First, even though we investigated the collision process using various methods,
the reason why such collisions occurs is not disclosed.
In the single $\phi^4$ model, kink dynamics is well explained by the collective coordinate method \cite{sugiyama1979kink, goodman2007chaotic, takyi2016collective}.
Therefore, we expect that the collective-coordinate method can explain multi-kink dynamics in the multifield nonlinear models.
Second, it would be interesting to study multi-kink collisions in multifield models, including higher-order terms such as $\phi^6$ and $\phi^8$ because interesting features (absence of internal modes, high energy density spot, etc.) are reported in the single-field models \cite{dorey2011kink,marjaneh2017multi,saxena2019higher}.

%%%%%%%%%%%%%%%%%%%%%%%%%%%%%%%%%%%%%%%%%%%%%%%%%%%%%%%%%%%%%%%%%%
%%%% Figure 11 F and V                                   %%%%%%%%%
%%%%%%%%%%%%%%%%%%%%%%%%%%%%%%%%%%%%%%%%%%%%%%%%%%%%%%%%%%%%%%%%%%
\begin{figure}[tbp]
\centering
\includegraphics[width=.8\textwidth,origin=c]{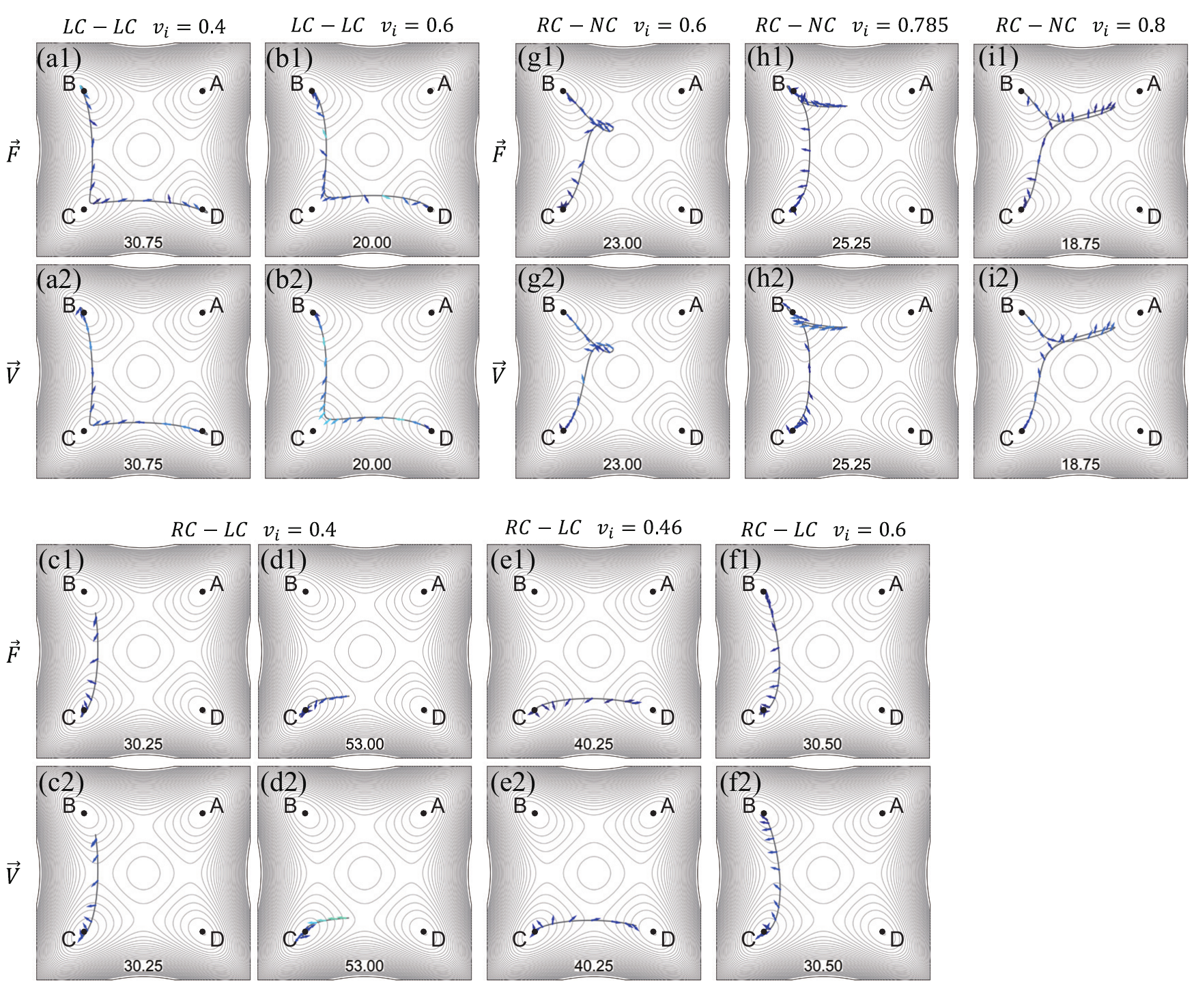}
\caption{
    \label{fig11:Force and Velocity} 
    % \textbf{Force and velocity vector fields in the field-space.}
    \textbf{The force $\vec F$ and velocity $\vec V$ vector fields for (a,b) LC-LC, (c-e) RC-LC, (g-i) RC-NC kink collisions.}
    %  for various initial velocities $v_i$.}
    The number in the bottom of each panel indicates the time of each snapshot.
    On the top of each plot, the initial velocities are indicated,
    which are the same velocities in figures~\ref{fig4:LC-LC}, \ref{fig5:LC LC Contours},
    \ref{fig6:RC-LC}, \ref{fig7:RC LC Contours}, \ref{fig8:RC-NC} and \ref{fig9:RC NC Contours}.
    The complete time-dependent force and velocity vector fields can be seen in the supplementary movie online.
    }
\end{figure}

\appendix
\section{Appendix}
\subsection{Coupled equations for internal modes for RC and LC kinks\label{app:coupled-equations}}
For RC and LC kinks, we consider the following small oscillatory modes ($\eta_{p}^{C}$ and $\eta_{l}^{C}$) around the static solution:
\begin{align}
    \label{a-eq:2.31:RC_internal_mode_1}
    &\phi_{i}(t,x) = \mu \tanh (\beta x) + \eta_{p}^{C}(x) e^{- i \omega t},
    \\
    \label{a-eq:2.31:RC_internal_mode_2}
    &\phi_{j}(t,x) = \mu - h(x) + \eta_{l}^{C}(x) e^{- i \omega t},
\end{align}
where $\beta=Z_1 m/\sqrt{2}$, $j=3-i$ ($i=1,2$), and $h(x)= C_1 \sech^2 \left( \frac{C_2 }{\sqrt{2}} x \right)$.
Using the field equation (\ref{eq:2.8:EoM}) and ignoring higher-order terms, the following linearized equations can be seen:
\begin{align*}
    & \left[ \frac{\partial^2 }{\partial t^2} - \frac{\partial^2 }{\partial x^2}
    - m^2 
    + 3 \lambda \mu^2 \tanh^2 (\beta x)
    + \alpha (\mu-h)^2
    \right] \eta_{p}^{C}
    + 2 \alpha \mu \tanh (\beta x) (\mu-h) \eta_{l}^{C}
    = 0,
    \\
    & \left[ \frac{\partial^2 }{\partial t^2} - \frac{\partial^2 }{\partial x^2}
    - m^2 
    + 3 \lambda (\mu - h)^2 + \alpha \mu^2 \tanh^2 (\beta x)
    \right] \eta_{l}^{C}
    + 2 \alpha \mu \tanh (\beta x) (\mu-h) \eta_{p}^{C}
    = 0.
\end{align*}
These coupled differential equations do not have analytic solutions.
However, in the leading order, we consider two fluctuations $\eta_{p}^{C}$ and $\eta_{l}^{C}$ separately and obtain the internal modes numerically.

\subsection{Force and velocity vector fields in the field space\label{app:force-velocity}}
In this appendix, we consider force and velocity vector fields ($\vec F$ and $\vec V$) of colliding kinks in field space.
The definitions for force and velocity vectors are given by (\ref{eq:3.4:force-velocity}).

Let's consider two collisions that have similar field-space orbits but different real-space field profiles.
In this case, the corresponding two $\vec T$ can be different because each $\vec T$ depends on the details of real-space field profiles.
On the other hand, the corresponding two $P_i$ are the same if two field-space orbits are the same.
Therefore, even though two collisions are very similar until a particular time and their field-space orbits are similar, the directions of $\vec T$ and $\vec V$ can be dramatically different between similar orbits. 
This plays an important role in determining the final collision results and understanding the collision processes.

For this, we compare many collisions that have very similar orbits but different collision results.
Figure~\ref{fig11:Force and Velocity} shows such orbits, $\vec F$, and $\vec V$ for LC-LC, RC-LC, and RC-NC kink collisions for various initial velocities for some particular moments.
The complete time evaluations can be seen in the supplementary movie online. 

First, we consider the LC-LC kink collisions.
Figure~\ref{fig11:Force and Velocity}(a) and figure~\ref{fig11:Force and Velocity}(b) show snapshots of the two collisions for $v_i=0.4$ and $v_i=0.6$, respectively. They correspond to figure~\ref{fig5:LC LC Contours}(b4) and figure~\ref{fig5:LC LC Contours}(d3), respectively.
In figure~\ref{fig11:Force and Velocity}(a,b), the shapes of each orbit are very similar but the directions of $\vec F$ and $\vec V$ are different between the two cases ($v_i=0.4$ and $v_i=0.6$).
Thus, in figure~\ref{fig11:Force and Velocity}(a), the orbit tries to become the $B\rightarrow C$ line due to $\vec F$ and $\vec V$.
On the other hand, in figure~\ref{fig11:Force and Velocity}(b), the orbit is dragged to the $C$ groundstate due to $\vec F$ and $\vec V$.
Note that even though the two orbits in figure~\ref{fig11:Force and Velocity}(a1,b1) are very similar, the two $\vec T$ (and hence $\vec F$) are different.
Therefore, the collision results diverge, as discussed in section~\ref{sec:LC-LC-RC-RC}.
As similar explanations are possible for the RC-LC and RC-NC kink collisions, they will be briefly discussed.

For the RC-LC kink collisions, figure~\ref{fig11:Force and Velocity}(c) and figure~\ref{fig11:Force and Velocity}(f) correspond to figure~\ref{fig7:RC LC Contours}(b3) and figure~\ref{fig7:RC LC Contours}(f3), respectively. In figure~\ref{fig11:Force and Velocity}(c,f), the shapes of each orbit are very similar but $\vec F$ and $\vec V$ are different. $\vec F$ and $\vec V$ drag the orbit toward the $C$ groundstate in figure~\ref{fig11:Force and Velocity}(c), while they pull the orbit toward the $B$ groundstate in figure~\ref{fig11:Force and Velocity}(f). Thus, the collision results are different for these two cases.
Similarly, the orbits in figure~\ref{fig11:Force and Velocity}(d,e) are alike, but $\vec F$ and $\vec V$ are different. $\vec F$ and $\vec V$ drag the orbit toward the $C$ groundstate in figure~\ref{fig11:Force and Velocity}(d), while they pull the orbit toward the $D$ groundstate in figure~\ref{fig11:Force and Velocity}(e).

For the RC-NC kink collisions, figure~\ref{fig11:Force and Velocity}(g) and figure~\ref{fig11:Force and Velocity}(i) correspond to figure~\ref{fig9:RC NC Contours}(b3) and figure~\ref{fig9:RC NC Contours}(f3), respectively. In figure~\ref{fig11:Force and Velocity}(g), $\vec F$ and $\vec V$ drag the orbit toward the $B \rightarrow C$ line, while they pull the orbit toward the $A$ groundstate in figure~\ref{fig11:Force and Velocity}(i). % Thus, the collision results are different for these two cases.
Similarly, the orbits in figure~\ref{fig11:Force and Velocity}(g,h) are alike, but $\vec F$ and $\vec V$ are different. $\vec F$ and $\vec V$ drag the orbit toward the $B \rightarrow C$ line in figure~\ref{fig11:Force and Velocity}(g), while they pull the orbit toward the $B$ groundstate in figure~\ref{fig11:Force and Velocity}(h).
Note that figure~\ref{fig11:Force and Velocity}(h) corresponds to figure~\ref{fig9:RC NC Contours}(d3).

\begin{acknowledgments}
We thank Tae-Hwan Kim for useful discussions.
J.-H.C., S.-H.H., M.K. and S.C. were supported by National  Research  Foundation  (NRF)  of Korea  through  Basic  Science  Research  Programs (NRF-2018R1C1B6007607, NRF-2021R1H1A1013517), the research fund of Hanyang University (HY-2017), and the POSCO Science Fellowship of POSCO TJ Park Foundation.
\end{acknowledgments}

\paragraph{Note added.} This is also a good position for notes added
after the paper has been written.

\bibliographystyle{JHEP}
\bibliography{reference.bib}

\end{document}